\documentclass[12pt]{iopart}

\pdfoutput=1

\usepackage{graphicx,sidecap}
\usepackage{bm}
\usepackage{braket}
\usepackage{amssymb}
\usepackage{cite}
\usepackage{mathrsfs}
\usepackage{amsmath}
\usepackage[dvipsnames]{xcolor}
\usepackage{hyperref}
\usepackage{enumerate}
\usepackage[toc,title]{appendix}
\usepackage{verbatim}
\usepackage[margin=30pt, bf, font=footnotesize, center, justification=justified]{caption}[2004/07/16]

\hypersetup{colorlinks,bookmarksopen,bookmarksnumbered,
citecolor=red,
linkcolor=blue,
pdfstartview=green,
urlcolor=teal}

\catcode`@=11 % If we need private TeX macros
\renewcommand\tableofcontents{%
  \section*{\contentsname}%
  \@starttoc{toc}%
}
\catcode`@=12% '@' is no more a character

\def\be{\begin{equation}}
\def\ee{\end{equation}}

\def\bea{\begin{eqnarray}}
\def\eea{\end{eqnarray}}

\begin{document}

\title[On entanglement hamiltonians of an interval in massless harmonic chains
]
{On entanglement hamiltonians of an interval \\ in massless harmonic chains
}

\vspace{.5cm}

\author{Giuseppe Di Giulio and Erik Tonni}
\address{SISSA and INFN Sezione di Trieste, via Bonomea 265, 34136 Trieste, Italy.}

\vspace{.5cm}

\begin{abstract}
We study the continuum limit of the entanglement hamiltonians 
of a block of consecutive sites in massless harmonic chains.
This block is either in the chain on the infinite line 
or at the beginning of a chain on the semi-infinite line with Dirichlet boundary conditions imposed at its origin. 
The entanglement hamiltonians of the interval predicted by Conformal Field Theory 
 for the massless scalar field are obtained in the continuum limit.
We also study the corresponding entanglement spectra 
and the numerical results for the ratios of the gaps are compatible with the operator content 
of the Boundary Conformal Field Theory of a massless scalar field with Neumann boundary conditions imposed 
along the boundaries introduced around the entangling points by the regularisation procedure.
\end{abstract}

\maketitle

\newpage

\tableofcontents

%%%%%%%%%%%%%%%%%%%%%%%%%%%%%%%%%%%%%%%%%%%%%%%%%%%%
%\newpage

\section{Introduction}
\label{sec:intro}

Entanglement has attracted an intense research activity during the last two decades,
mostly focused on theoretical approaches \cite{ep-rev,ch-rev, other-rev, holog-rev},
but in the last few years also experimental setups have been realised to detect its
characteristic features \cite{experiments}.

Given a quantum system in a state described by the density matrix $\rho$,
assuming that its Hilbert space can be factorised as $\mathcal{H} = \mathcal{H}_A \otimes \mathcal{H}_B$, 
the reduced density matrix $\rho_A$ is defined
by tracing out the degrees of freedom of $\mathcal{H}_B$,
namely by $\rho_A = \textrm{Tr}_{\mathcal{H}_B} \rho$,
with the normalisation condition $\textrm{Tr}_{\mathcal{H}_A} \rho_A =1$.
%%%
The reduced density matrix can be written as  $\rho_A = e^{-K_A}/\mathcal{Z}_A$, 
where the hermitian operator $K_A$ is the  entanglement hamiltonian (also known as modular hamiltonian)
and $\mathcal{Z}_A = \textrm{Tr}_{\mathcal{H}_A} \,e^{-K_A}$.
The entanglement entropy is easily obtained from the eigenvalues of $\rho_A$
\cite{ee-initial-papers, wil-larsen-94, cc-04}.
Important results have been obtained for factorisations of the Hilbert space
corresponding to bipartitions $A \cup B$ of the space, 
namely when $A$ is a spatial region and $B$ its complement.
In these cases the hypersurface $\partial A = \partial B$ separating $A$ and $B$ is called entangling hypersurface.

A fundamental theorem proved by Bisognano and Wichmann \cite{bw}
in the context of Algebraic Quantum Field Theory claims that,
given a relativistic Quantum Field Theory (QFT) in $d+1$ dimensions in its vacuum state
(we denote by $\boldsymbol{x}$ the $d$ dimensional position vector)
and the spatial bipartition where $A$ corresponds to half space
and the entangling hypersurface is the flat hyperplane,
the entanglement hamiltonian of $A$ can be written as an integral over the half space $A$
of the energy density $T_{00}(\boldsymbol{x})$ of the QFT
as follows
\be
\label{EH-BW-intro}
K_A 
\,=\, 2\pi \int_A \! x_1 \,T_{00}(\boldsymbol{x}) \, d\boldsymbol{x}\,.
\ee

When the QFT is a Conformal Field Theory (CFT), 
the conformal symmetry allows
to write analytic expressions for the entanglement hamiltonians for simple spatial bipartitions, 
mainly at equilibrium \cite{hislop-longo, chm, klich-13, ct-16, trs-18-rainbow}
but in few cases also out of equilibrium \cite{ct-16}.
More complicated bipartitions require a detailed knowledge of the underlying CFT
\cite{ch-09-eh-2int, Arias-18}.

\begin{figure}[t!]
\vspace{-.2cm}
\hspace{-0.cm}
\begin{center}
\includegraphics[width=1\textwidth]{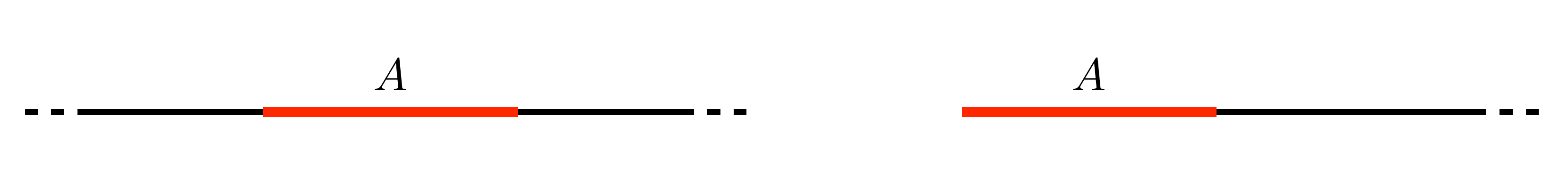}
\end{center}
\vspace{-.3cm}
\caption{
The spatial bipartitions considered in this manuscript:
an interval $A$ in the infinite line (left)
and an interval $A$ at the beginning of the semi-infinite line (right).
}
\label{fig:config}
\end{figure}

In a $1+1$ dimensional  CFT at equilibrium in its vacuum state,
we consider bipartitions where $A$ is an interval and such that its entanglement hamiltonian can be written as
\be
\label{EH-intro}
K_A 
\,=\, \ell \int_A \beta(x) \, T_{00}(x) \, dx
\ee
where $\ell$ is the length of $A$ and $\beta(x)$ depends on the bipartition. 
In this manuscript we focus on the bipartitions shown in Fig.\,\ref{fig:config}.
%%%
For an interval in the infinite line (left panel in Fig.\,\ref{fig:config}),
the weight function in (\ref{EH-intro})  is the following parabola \cite{hislop-longo, chm}
\be
\label{parabola-CFT-infinite}
\beta(x)=2\pi\, \frac{x}{\ell} \left( 1-  \frac{x}{\ell}\, \right).
\ee
When $A$ is an interval at the beginning of a semi-infinite line 
(right panel in Fig.\,\ref{fig:config}), 
the weight function in (\ref{EH-intro}) is the half parabola given by \cite{ct-16}
\be
\label{parabola-CFT-semi-infinite}
\beta(x)= \pi \left(\, \frac{x}{\ell}+1 \right) \left( 1-  \frac{x}{\ell}\, \right)
\ee
independently of the boundary conditions imposed at the beginning of the semi-infinite line.
%%%
It is interesting to explore the procedure that allows to obtain these 
entanglement hamiltonians in CFT as the continuum limit
of the corresponding entanglement hamiltonians in the lattice models
providing a discretisation of the CFT.

%%%
Entanglement hamiltonians in free lattice models 
at equilibrium in their ground state have been studied in
\cite{ep-rev, ch-rev, Peschel-eh-free-models, peschel-04-eh-fermions, ep-17, ep-18, Arias-16, etp-19}.
A detailed analysis of the continuum limit has been recently carried out 
for an interval in an infinite chain of free fermions \cite{etp-19},
by employing the analytic results obtained by Eisler and Peschel in \cite{ep-17}.
%%%
In this manuscript we study the continuum limit of the entanglement hamiltonians 
of a block of consecutive sites in massless harmonic chains
by following the approach of \cite{etp-19},
which is based on the observation that, in this limit,  
the proper combinations of all the diagonals of the matrices 
determining the entanglement hamiltonian on the lattice must be considered \cite{Arias-16}.

The eigenvalues $\lambda_j$ of the entanglement hamiltonian provide the entanglement spectrum,
which contains relevant physical information \cite{ent-spectrum-top}.
It is worth introducing the gaps $g_r \equiv \log \lambda_{\textrm{\tiny max}} - \log \lambda_r$
with respect to the largest eigenvalue
and also their ratios $g_r / g_1$ with respect to the smallest gap $g_1$.
We remark that these ratios are not influenced 
by a global shift and a rescaling of the entire spectrum.

In a two dimensional QFT in imaginary time,
a useful way to regularise the ultraviolet (UV) divergences
consists in removing infinitesimal disks whose radius is the UV cutoff
around the entangling points of the bipartition
\cite{wil-larsen-94, peschel-04-eh-fermions, tachikawa, ct-16}.
In two dimensional CFT, this regularisation procedure leads to a Boundary Conformal Field Theory (BCFT) \cite{bcft}
if proper conformal boundary conditions are imposed along the boundaries in the euclidean spacetime
(both the boundaries given by the physical boundaries of the system and 
the ones due to this regularisation procedure must be considered). 
For a class of entanglement hamiltonians which includes the ones we are interested in, 
it has been found that \cite{ct-16}
\be
\label{ratios_bcft_intro}
\frac{g_r}{g_1} = \frac{\Delta_r}{\Delta_1}
\ee
where $r\geqslant 1$ and $\Delta_r >0$ are the non vanishing elements of the conformal spectrum
(made by conformal dimensions of the primary fields and of their descendants)
of the underlying BCFT.
Numerical evidences that the conformal spectrum of a BCFT provides the entanglement spectrum 
have been first obtained at equilibrium by L\"auchli in \cite{lauchli-spectrum}
and more recently also out of equilibrium 
\cite{dat-19, stt-19}.

In this manuscript we focus on massless harmonic chains
and perform a numerical analysis of the continuum limit of 
two entanglement hamiltonians of an interval $A$
and of the corresponding entanglement spectra.
We consider a massless harmonic chain
both on the infinite line 
and on the semi-infinite line with Dirichlet boundary conditions imposed at its origin.
The continuum limit of these lattice models is the CFT given by the massless scalar field $\Phi$, 
whose central charge is $c=1$.
By introducing the canonical momentum field $\Pi = -\,  \partial_t \Phi$,
the energy density on the infinite line reads
\be
\label{T00 massless infinite}
T_{00}(x) = 
\frac{1}{2} \, \Big[\,
\Pi(x)^2 + \big( \partial_x \Phi(x)\big)^2 
\, \Big]
\ee
and on the semi-infinite line is given by \cite{mintchev-liguori}
\be
\label{T00 massless semi-infinite}
T_{00}(x) = 
\frac{1}{2} \, \Big[\,
\Pi(x)^2 - \Phi(x)\, \partial^2_x \Phi(x)
\, \Big]\,.
\ee

We study the spatial bipartitions shown in Fig.\,\ref{fig:config},
whose entanglement hamiltonians predicted by CFT are given by
(\ref{EH-intro}), (\ref{parabola-CFT-infinite}) and (\ref{T00 massless infinite})
for the interval in the infinite line
and by (\ref{EH-intro}), (\ref{parabola-CFT-semi-infinite}) and (\ref{T00 massless semi-infinite})
for the interval at the beginning of the semi-infinite line.
Our numerical analysis is based on the procedure described in \cite{ep-17, etp-19} 
to study the continuum limit of the entanglement hamiltonian of an interval in an infinite chain of free fermions.
We study also the entanglement spectra of these entanglement hamiltonians, 
finding that the CFT prediction (\ref{ratios_bcft_intro}) holds, once Neumann boundary conditions
are imposed along the boundaries introduced by the regularisation procedure.

This manuscript is organised as follows. 
In \S\ref{sec:EHinHC} we report  the entanglement hamiltonian 
of an interval in harmonic chains in terms of the two-point correlators.
In \S\ref{sec:infinite-line} we study the continuum limit of the 
entanglement hamiltonian of an interval in the infinite line and 
in \S\ref{sec:semi-infinite-line} this analysis is performed for 
an interval at the beginning of the semi-infinite line with 
Dirichlet boundary conditions. 
In \S\ref{sec:conclusions} we draw some conclusions.
The Appendix\;\ref{sec_app:details} contains 
further results supporting some observations made in the main text.

%\newpage
%%%%%%%%%%%%%%%%%%%%%%%%%%%%%%%%%%%%%%%%%%%%%%%%%%%%%%%%
%%%%%%%%%%%%%%%%%%%%%%%%%%%%%%%%%%%%%%%%%%%%%%%%%%%%%%%%
\section{Entanglement hamiltonians in the harmonic chain}
\label{sec:EHinHC}

In this section we report the expression of the entanglement hamiltonian of a subsystem
in harmonic chains in terms of the two-point correlators \cite{ch-rev},
discussing also some decompositions that will be employed throughout the manuscript.

The hamiltonian of the harmonic chain
with nearest neighbour spring-like interaction reads
\be
\label{HC ham}
\widehat{H} = \sum_{i} \left(
\frac{1}{2m}\,\hat{p}_i^2+\frac{m\omega^2}{2}\,\hat{q}_i^2 +\frac{\kappa}{2}(\hat{q}_{i+1} -\hat{q}_i)^2
\right)
\ee
where the position and the momentum operators  $\hat{q}_i$ and $\hat{p}_i$
are hermitean operators satisfying the canonical commutation relations
$[\hat{q}_i , \hat{q}_j]=[\hat{p}_i , \hat{p}_j] = 0$ 
and $[\hat{q}_i , \hat{q}_j]= \textrm{i} \delta_{i,j}$
($\hbar =1$ throughout this manuscript).
In our numerical analysis we set $\kappa=m=1$.

The hamiltonian (\ref{HC ham}) is the discretization of the hamiltonian of a massive scalar field in the continuum,
whose massless regime given by $\omega =0$
is a CFT with central charge $c=1$.
The range of the index $i$ in (\ref{HC ham}) depends on the 
spatial domain supporting the harmonic chain:
in this manuscript we consider either the infinite line ($i \in \mathbb{Z}$)
or the semi-infinite line (integer $i \geqslant 0$). 
When (\ref{HC ham}) is defined on the semi-infinite line, 
it is crucial to specify also the boundary condition imposed at the beginning of the semi-infinite line (i.e. at $i=0$)
and in our analysis we consider the case of Dirichlet boundary conditions. 
The two-point correlators $Q_{ij} =  \langle \hat{q}_i \,\hat{q}_j \rangle$
and $P_{ij} = \langle \hat{p}_i \,\hat{p}_j \rangle$ 
provide the generic elements of the correlation matrices $Q$ and $P$ 
respectively.

Let us consider harmonic chains (\ref{HC ham}) in their ground state $| 0 \rangle$
and introduce the bipartition of the chain into a spatial domain $A$ made by $L$ sites and its complement $B$,
assuming that the Hilbert space can be bipartite accordingly as $\mathcal{H} = \mathcal{H}_A \otimes \mathcal{H}_B$.
Since for these quantum systems
the reduced density matrix $\rho_A \equiv \textrm{Tr}_{\mathcal{H}_B} (| 0 \rangle \langle 0 |)$ 
remains Gaussian, independently of the choice of the bipartition, 
the corresponding entanglement hamiltonian $\widehat{K}_A$ 
is a quadratic hermitian  operator, which can be written as follows
\cite{Peschel-eh-free-models, ch-rev}
\be
\label{ent-ham HC}
\widehat{K}_A
= \frac{1}{2}\, \boldsymbol{\hat{r}}^{\textrm t} H_A\, \boldsymbol{\hat{r}}
\hspace{.5cm} \qquad \hspace{.5cm} 
\boldsymbol{\hat{r}} = 
\bigg( \hspace{-.05cm} 
\begin{array}{c}
\boldsymbol{\hat q} \\  \boldsymbol{\hat p}
\end{array} \hspace{-.05cm}  \bigg)
\ee
where the $2L$ dimensional vector $\boldsymbol{\hat{r}}$ collects the position and the momentum operators 
$\hat{q}_i$ and $\hat{p}_i$ with $i \in A$.
The matrix $H_A$
is real, symmetric and positive definite; 
hence $\widehat{K}_A$ is hermitian.
%%%%%
In terms of the reduced correlation matrices $Q_A$ and $P_A$,
obtained by restricting $Q$ and $P$ to the subsystem $A$,
the matrix $H_A$ can be evaluated as follows \cite{ch-rev}
\bea
\label{eh-block-ch-version}
H_A 
= 
M \oplus N
&\equiv &
\Big( h\big(\sqrt{P_A Q_A}\,\big)   \oplus h\big(\sqrt{Q_A P_A}\,\big) \Big) 
 \big( P_A \oplus Q_A \big)
 \\
&=&
 \big( P_A \oplus Q_A \big)
 \Big( h\big(\sqrt{Q_A P_A}\,\big)   \oplus h\big(\sqrt{P_A Q_A}\,\big) \Big) 
 \nonumber
\eea
where 
\be
\label{h-function def}
h(y)  \equiv  \frac{1}{y} \, \log \!\left(\frac{y +1/2}{y - 1/2} \right).
\ee
The equivalence of the two expressions in (\ref{eh-block-ch-version}) 
can be verified by transposing one of them
and employing that $M$ and $N$ are symmetric 
(we also need $(\sqrt{Q_A P_A}\,)^{\textrm t} = \sqrt{P_A Q_A}$, 
which is easily obtained from the fact that $Q$ and $P$ are symmetric).

In this manuscript we study entanglement hamiltonians $\widehat{K}_A$
when the entire chain is at equilibrium in its ground state
and when the subsystem $A$ is a block of $L$ consecutive sites
either in the infinite line 
or at the beginning of the semi-infinite line 
where Dirichlet boundary conditions are imposed
(see Fig.\,\ref{fig:config}). 
In these two cases, the matrix $H_A$ is block diagonal.
The off-diagonal blocks of $H_A$ 
can be non vanishing e.g. for the time dependent entanglement hamiltonians
after a global quantum quench \cite{dat-19}.

The matrix $H_A$ can be constructed numerically through 
(\ref{eh-block-ch-version}) and (\ref{h-function def}).
In order to employ these formulas, the eigenvalues of 
the matrix $\sqrt{Q_A P_A}$ must be strictly larger than $1/2$.
In our numerical analysis many eigenvalues very close to $1/2$ occur
and the software automatically approximates them to $1/2$ 
whenever a low numerical precision is set throughout
the numerical analysis. 
This forces us to work with very high numerical precisions. 
In particular
we have employed precisions up to 6500 digits,
depending on the specific calculation. 
We observe that 
higher precision is required as
$L$ and $\omega$ increase.

The expressions (\ref{ent-ham HC}) and (\ref{eh-block-ch-version}) naturally lead to write  
the entanglement hamiltonian 
in terms of the symmetric matrices $M$ and $N$ as follows
\be
\label{KA/2}
\widehat{K}_A = \frac{\widehat{H}_M + \widehat{H}_N }{2} 
\ee
where
\be
\label{H_M and H_N operators}
\widehat{H}_M
\equiv
\sum_{i,j =1}^L M_{i,j} \,\hat{q}_i \, \hat{q}_j
\;\;\qquad\;\;
\widehat{H}_N
\equiv
\sum_{i,j =1}^L N_{i,j} \,\hat{p}_i \, \hat{p}_j\,.
\ee

These sums can be organised in different ways.
For instance, by writing the symmetric matrices  $M$ and $N$ 
as sums of a diagonal matrix, an upper triangular matrix and a lower triangular matrix,
it is straightforward to obtain
\bea
\label{H_M asymmetric}
\widehat{H}_M
&=&
L
\sum_{i=1}^L 
\bigg( \frac{M_{i,i}}{L} \,\hat{q}_i^2
+
2 \sum_{k =1}^{L-i} \frac{M_{i,i+k}}{L} \;\hat{q}_i \, \hat{q}_{i+k}
\bigg)
\\
\label{H_N asymmetric}
\rule{0pt}{.9cm}
\widehat{H}_N
&=&
L
\sum_{i=1}^L 
\bigg( \frac{N_{i,i}}{L} \,\hat{p}_i^2
+
2 \sum_{k =1}^{L-i} \frac{N_{i,i+k}}{L} \;\hat{p}_i \, \hat{p}_{i+k}
\bigg)\,.
\eea

In \cite{Arias-16} the sums (\ref{H_M and H_N operators}) have been rewritten 
by decomposing the contribution coming from the $i$-th row of the matrices $M$ and $N$,
and this leads to
\bea
\label{H_M Casini}
\widehat{H}_M
&=&
L
\sum_{i=1}^L 
\bigg( \frac{M_{i,i}}{L} \,\hat{q}_i^2
+
 \sum_{k =1}^{L-i} \frac{M_{i,i+k}}{L} \;\hat{q}_i \, \hat{q}_{i+k}
 +
\sum_{k =1}^{i-1} \frac{M_{i,i-k}}{L} \;\hat{q}_i \, \hat{q}_{i-k}
\bigg)
\\
\label{H_N Casini}
\widehat{H}_N
&=&
L
\sum_{i=1}^L 
\bigg( \frac{N_{i,i}}{L} \,\hat{p}_i^2
+
 \sum_{k =1}^{L-i} \frac{N_{i,i+k}}{L} \;\hat{p}_i \, \hat{p}_{i+k}
 +
\sum_{k =1}^{i-1} \frac{N_{i,i-k}}{L} \;\hat{p}_i \, \hat{p}_{i-k}
\bigg)\,.
\eea

We find it convenient to introduce also the following decomposition
\bea
\label{H_M symm}
& & \hspace{-2.5cm}
\widehat{H}_M
=
L\,
\Bigg[\,
\sum_{i=1}^{L/2}
\bigg( \frac{M_{i,i}}{L} \,\hat{q}_i^2
+ \!\!
\sum_{k=1}^{L-2i+1}
\frac{M_{i,i+k}}{L} \;\hat{q}_i \, \hat{q}_{i+k}
\bigg)
+\!\!\!
\sum_{i=L/2+1}^{L} \!\!
\bigg( \frac{M_{i,i}}{L} \,\hat{q}_i^2
+\!\!
\sum_{k=1}^{2i-L-2}
\frac{M_{i-k,i}}{L} \;\hat{q}_i \, \hat{q}_{i-k}
\bigg)
\Bigg]
\nonumber\\
& &
\\
\label{H_N symm}
& & \hspace{-2.5cm}
\widehat{H}_N
=
L\,
\Bigg[\,
\sum_{i=1}^{L/2}
\bigg( \frac{N_{i,i}}{L} \,\hat{p}_i^2
+ \!\!
\sum_{k=1}^{L-2i+1}
\frac{N_{i,i+k}}{L} \;\hat{p}_i \, \hat{p}_{i+k}
\bigg)
+\!\!\!
\sum_{i=L/2+1}^{L}
\bigg( \frac{N_{i,i}}{L} \,\hat{p}_i^2
+\!\!
\sum_{k=1}^{2i-L-2}
\frac{N_{i-k,i}}{L} \;\hat{p}_i \, \hat{p}_{i-k}
\bigg)
\Bigg]
\nonumber\\
& &
\eea
where the contribution of the counter diagonal has been included into the summation over $1\leqslant i \leqslant L/2$.
This choice leads to an inconsistency in the range of $k$ of the last double sum when $i=L/2+1$, 
which can be easily fixed by imposing the vanishing of this term.

An alternative decomposition, inspired by the numerical analysis performed in \cite{etp-19},
is discussed in Appendix\;\ref{sec_app:details}.
In our numerical analysis we have tested all the decompositions introduced above and in Appendix\;\ref{sec_app:details}
for both the spatial bipartitions shown in Fig.\,\ref{fig:config}.
In the main text we show the numerical results obtained from (\ref{H_M symm}) and (\ref{H_N symm}), 
while the results found through the other decompositions
have been reported in Appendix\;\ref{sec_app:details}.

The analytic results for the entanglement hamiltonian of an interval in the infinite chain of free fermions
found in \cite{ep-17, etp-19} 
and the decompositions introduced above for the operators $\widehat{H}_M$ and $\widehat{H}_N$
suggest to introduce the following limits
\be
\label{munuk def}
\lim_{L \to \infty} \, \frac{M_{i,i+k}}{L}  \,\equiv\, \mu_k(x_k)
\;\;\qquad\;\;
\lim_{L \to \infty} \, \frac{N_{i,i+k}}{L}  \,\equiv\, \nu_k(x_k)
\hspace{.5cm} \qquad \hspace{.5cm}
x_k 
\equiv \frac{1}{L} \left( i +\frac{k}{2}\right)
\ee
where $i + k/2$ is the midpoint between the $i$-th and the $(i+k)$-th site. 
The existence of the functions $\mu_k$ and $\nu_k$
is a crucial assumption in the subsequent 
derivations of the CFT predictions for the entanglement hamiltonians.

%\newpage
%%%%%%%%%%%%%%%%%%%%%%%%%%%%%%%%%%%%%%%%%%%%%%%%%%%%%%%%
%%%%%%%%%%%%%%%%%%%%%%%%%%%%%%%%%%%%%%%%%%%%%%%%%%%%%%%%
\section{Interval in the infinite line}
\label{sec:infinite-line}

In this section we consider the harmonic chain on the infinite line 
and perform a numerical analysis to study the continuum limit of the 
entanglement hamiltonian of an interval. 
We follow the procedure discussed in \cite{ep-17, etp-19}
for the continuum limit of the entanglement hamiltonian 
of an interval in the infinite chain of free fermions.
In \S\ref{sec:infinite-line-corr} we introduce the two-point correlators to construct $Q_A$ and $P_A$
and in \S\ref{sec:infinite-line-massless} we report the main analysis,
which leads to the CFT prediction (\ref{EH-intro}),
with the weight function (\ref{parabola-CFT-infinite}) and the energy density (\ref{T00 massless infinite}).
The entanglement spectrum is explored in \S\ref{sec:infinite-line-gaps}.

\subsection{Correlators}
\label{sec:infinite-line-corr}

The two-point correlators $ \langle \hat{q}_i \hat{q}_j  \rangle $ and $\langle \hat{p}_i \hat{p}_j  \rangle$ in the ground state 
of a finite harmonic chain made by $\mathcal{L}$ sites ($1\leqslant i \leqslant \mathcal{L}$ in (\ref{HC ham}))
with periodic boundary conditions ($\hat{q}_1 = \hat{q}_{\mathcal{L}+1}$ and $\hat{p}_1 = \hat{p}_{\mathcal{L}+1}$)
are respectively
 \be
 \label{corrs-periodic}
 \langle \hat{q}_i \hat{q}_j  \rangle =
\frac{1}{2 \mathcal{L}} \sum_{k=1}^{\mathcal{L}} \frac{1}{m \tilde{\omega}_k} \, \cos[2 \pi k\,(i-j)/\mathcal{L}] 
 \qquad
\langle \hat{p}_i \hat{p}_j  \rangle =
\frac{1}{2 \mathcal{L}} \sum_{k=1}^{\mathcal{L}} m \tilde{\omega}_k \, \cos[2 \pi k\,(i-j)/\mathcal{L}]
 \ee
where the dispersion relation reads
\be
\tilde{\omega}_k \equiv 
\sqrt{\omega^2 +\frac{4\kappa}{m}\, \big[ \sin(\pi k/\mathcal{L}) \big]^2} \,\geqslant \,\omega\,
\qquad
1 \leqslant k \leqslant \mathcal{L} \,.
\ee

The translation invariance induces the occurrence of the zero mode, that corresponds to $k = \mathcal{L}$.
Since $\tilde{\omega}_{\mathcal{L}} = \omega$,
it is straightforward to observe that $\langle \hat{q}_i \hat{q}_j  \rangle$ diverges as $\omega \to 0$;
hence the mass cannot be set to zero in the numerical analysis of this system. 

In the thermodynamic limit $\mathcal{L} \to + \infty$, the correlators in (\ref{corrs-periodic})  can be written as the integrals
\bea
\label{qq-corr-pbc}
\langle \hat{q}_i \hat{q}_j \rangle
&=&
\frac{1}{4 \pi\, m} \int_0^{2\pi} \!\!\!
\frac{\cos[\theta\, (i-j)] }{\sqrt{\omega^2+ (4\kappa/m)  \big[ \sin(\theta/2)\big]^2}} \;  d\theta
\\
\rule{0pt}{.9cm}
\label{pp-corr-pbc}
\langle \hat{p}_i \hat{p}_j \rangle
&=&
\frac{m}{4 \pi} \int_0^{2\pi}\!\!
\sqrt{\omega^2+ \frac{4\kappa}{m} \, \big[ \sin(\theta/2)\big]^2}
\; \cos[\theta\, (i-j)]\, d\theta
\eea
which can be evaluated analytically, finding the following expressions 
in terms of the hypergeometric function \cite{br-04}
\bea
\label{qq-corr-pbc-int}
& & \hspace{-1.5cm}
\langle \hat{q}_i \hat{q}_j \rangle
\,=\,
\frac{\zeta^{i-j+1/2}}{2\,\sqrt{\kappa m}}\,
\binom{i-j-1/2}{i-j}\,
_2F_1 \big(  1/2\, , i-j+ 1/2\, , i-j+1 \,,  \zeta^2 \,\big)
\\
\rule{0pt}{.9cm}
\label{pp-corr-pbc-int}
& & \hspace{-1.5cm}
\langle \hat{p}_i \hat{p}_j \rangle
\,=\,
\frac{\sqrt{\kappa m}\; \zeta^{i-j-1/2}}{2}\,
\binom{i-j-3/2}{i-j}\;
_2F_1 \big( - 1/2\, , i-j- 1/2\, , i-j+1 \,,  \zeta^2 \,\big)
\eea
where the parameter $\zeta$ is defined as
\be
\label{z-def}
\zeta \equiv \frac{\big(\omega\,-\, \sqrt{\omega^2+4\kappa/m} \,\big)^2}{4\kappa/m}\,.
\ee

The reduced correlation matrices $Q_A$ and $P_A$ are obtained 
by restricting  the indices $i$ and $j$ of the correlators (\ref{qq-corr-pbc-int}) and (\ref{pp-corr-pbc-int}) 
to the interval $A$, i.e. to the integer values in $[1, L]$.
By employing these reduced correlation matrices into (\ref{eh-block-ch-version}),
one finds the entanglement hamiltonian matrix $H_A$.
The entanglement hamiltonian of the interval in the infinite line is obtained
by plugging the matrix $H_A$ into (\ref{ent-ham HC}).

\subsection{Entanglement hamiltonian}
\label{sec:infinite-line-massless}

The entanglement hamiltonian of a block made by $L$ consecutive sites in the infinite line,
when the entire harmonic chain is in its ground state, 
is the operator constructed as explained in \S\ref{sec:infinite-line-corr}.
Considering the massless regime, we study 
the procedure to obtain the CFT prediction (\ref{EH-intro}),
with $\beta(x)$ and $T_{00}(x)$ given by (\ref{parabola-CFT-infinite})
and (\ref{T00 massless infinite}) respectively, 
through a numerical analysis of the continuum limit.

The translation invariance of the entire system prevents us from setting $\omega =0$ in our numerical analysis,
as already remarked in \S\ref{sec:infinite-line-corr}.
The data reported in all the figures discussed in this subsection have been obtained for $\omega L =10^{-500}$.
The choice of this value is discussed in Appendix\;\ref{sec_app:details}.

\begin{figure}[t!]
\vspace{.2cm}
\hspace{-.9cm}
% \begin{center}
\includegraphics[width=1.05\textwidth]{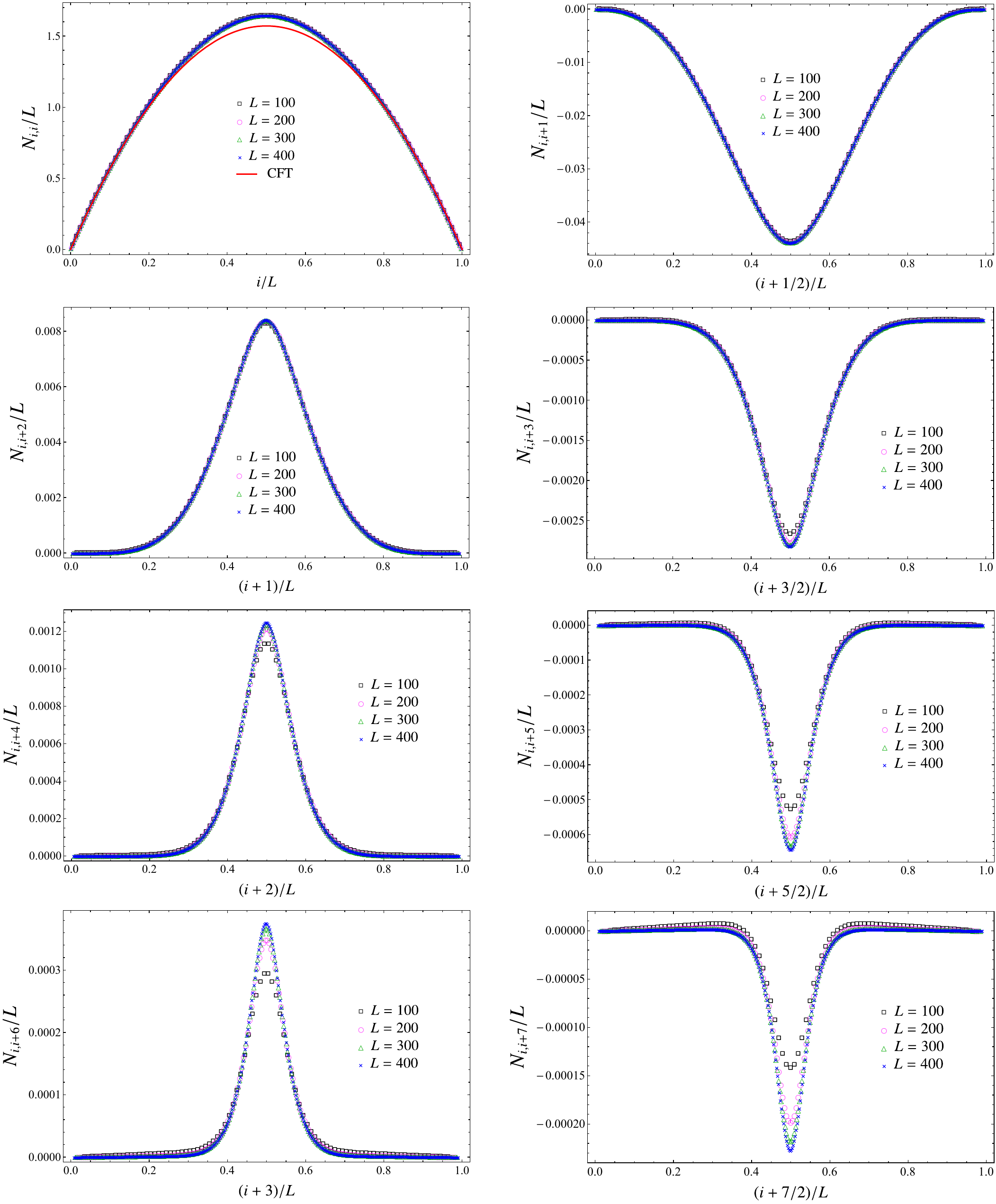}
% \end{center}
\vspace{-.5cm}
\caption{
Diagonals of the  matrix $N$ (see (\ref{munuk def}))
when the subsystem is an interval made by $L$ sites in the infinite line
and $\omega L =10^{-500}$.
The red solid curve is the parabola (\ref{parabola-CFT-infinite}).
}
\vspace{.5cm}
\label{fig:Nuk-Infinite}
\end{figure}

In Fig.\,\ref{fig:Muk-Infinite} and Fig.\,\ref{fig:Nuk-Infinite} we show the data 
for the diagonals $M_{i,i+k} / L$ and $N_{i,i+k} / L$ with $0 \leqslant k \leqslant 7$
for some values of $L$.
This numerical results lead to conclude that the limits in (\ref{munuk def})
provide well defined functions.
Furthermore, these functions have a well defined sign given by the parity of $k$,
are symmetric under reflection with respect to the center of the interval
(we checked numerically that this symmetry holds also for the data points, 
i.e. that $M_{i,i+p}=M_{L-i-p+1,L-i+1}$ and $N_{i,i+p}=N_{L-i-p+1,L-i+1}$)
and the absolute value of their maximum significantly decreases as $k$ increases. 
It would be interesting to find analytic expressions for the functions defined through the limits (\ref{munuk def}),
as done in \cite{ep-17} for the interval in the infinite chain of free fermions.

Assuming that  the functions $\mu_k$ and $\nu_k$ introduced in (\ref{munuk def}) are well defined, 
let us study the continuum limit of the entanglement hamiltonian (\ref{KA/2}),
where the quadratic operators $\widehat{H}_M$ and $\widehat{H}_N$
have been introduced in (\ref{H_M and H_N operators}).
In the following we adapt the procedure discussed 
in \cite{etp-19} for the entanglement hamiltonian of an interval in the infinite chain of free fermions. 
%%%
The quadratic operators $\widehat{H}_M$ and $\widehat{H}_N$
can be decomposed in different ways,
as discussed in \S\ref{sec:EHinHC} and in Appendix\;\ref{sec_app:details}.
For the sake of simplicity, in the following we describe the continuum limit for 
the decomposition given by 
(\ref{H_M asymmetric}) and (\ref{H_N asymmetric}),
but the procedure can be easily adapted to the ones given by 
(\ref{H_M Casini}) and (\ref{H_N Casini})
or by (\ref{H_M symm}) and (\ref{H_N symm}).
In Appendix\;\ref{sec_app:details} 
we discuss another decomposition, inspired by the numerical analysis performed in \cite{etp-19}.

\begin{figure}[t!]
\vspace{.2cm}
\hspace{-.9cm}
% \begin{center}
\includegraphics[width=1.05\textwidth]{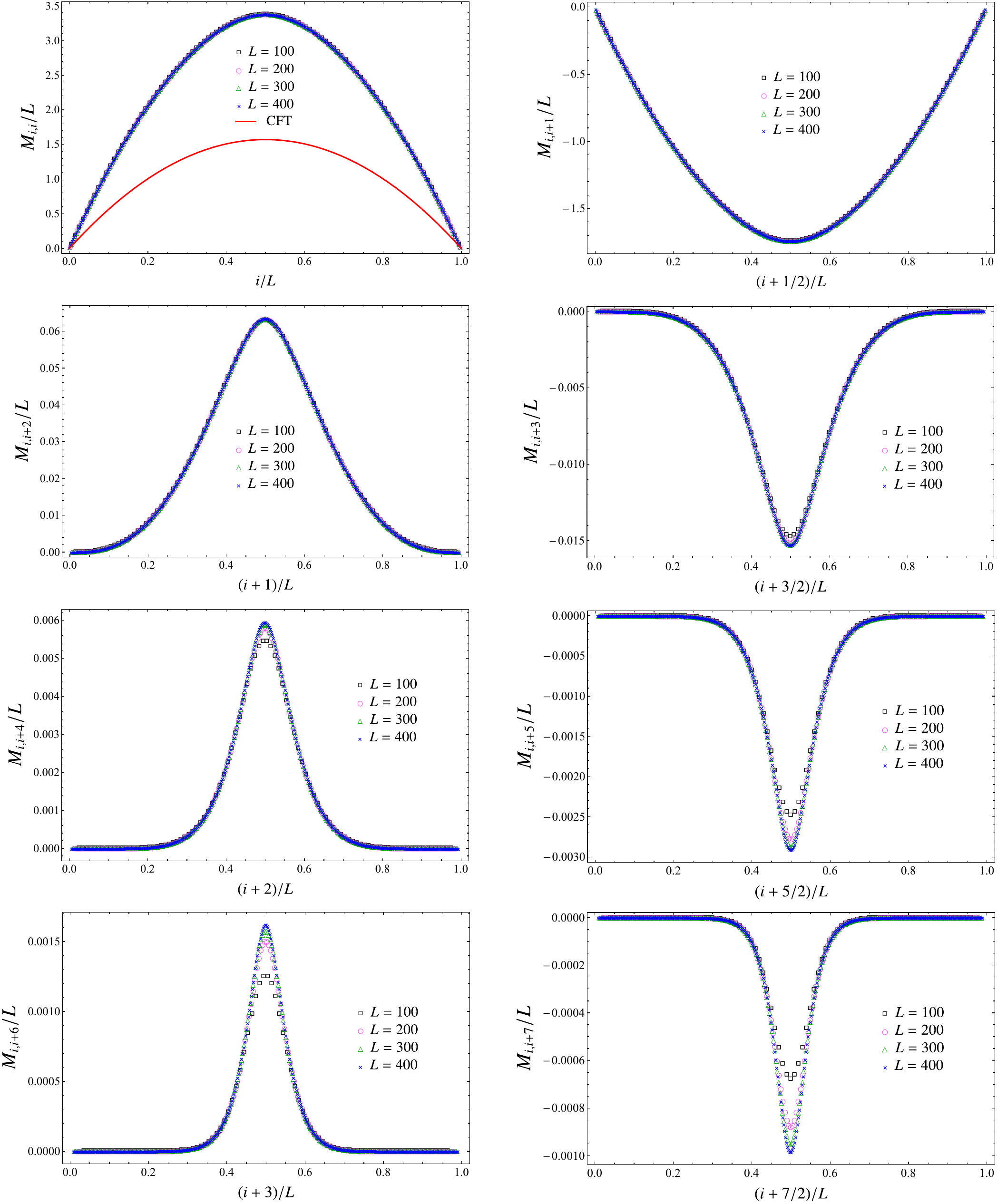}
% \end{center}
\vspace{-.5cm}
\caption{
Diagonals of the  matrix $M$ (see (\ref{munuk def}))
when the subsystem is an interval made by $L$ sites in the infinite line
and $\omega L =10^{-500}$.
The red solid curve is the parabola (\ref{parabola-CFT-infinite}).
}
\vspace{.5cm}
\label{fig:Muk-Infinite}
\end{figure}

The continuum limit is defined through the infinitesimal ultraviolet (UV) cutoff $a$:
it corresponds to take $a \to 0$ and $L \to \infty$  while $L a =\ell$ is kept constant. 
The position within the interval is labelled by $x = i a$ with $0< x<\ell$.
This leads to write the independent variable in (\ref{munuk def}) as follows
\be
x_k 
= \frac{x}{La} +\frac{k a/2}{La}
\ee
which tells that $ \mu_k =  \mu_k(x+ ka/2)$ and $ \nu_k =  \nu_k(x+ ka/2)$.

In the continuum limit, the fields $\Phi(x)$ and $\Pi(x)$ are introduced through 
the position and momentum operators as follows
\cite{altland simons book FT}
\be 
\label{qp-field-replacement}
\hat{q}_i \, \longrightarrow \,  \Phi(x)
\;\; \qquad \;\;
\hat{p}_i \, \longrightarrow \, a \, \Pi(x)
\ee
where the UV cutoff guarantees that these fields satisfy the
canonical commutation relations in the continuum limit, where the delta function occurs.
The operators $\hat{q}_{i+k} $ and $\hat{p}_{i+k} $ in (\ref{H_M asymmetric}) and (\ref{H_N asymmetric})
lead to fields whose argument is properly shifted. 
By employing (\ref{qp-field-replacement}) and the Taylor expansion as $a \to 0$,  
in the continuum limit it is straightforward to obtain that
\be
\label{q-der-field-replacement}
\hat{q}_{i+k} 
\; \longrightarrow \;
\Phi(x+ k a)
=
 \sum_{p \geqslant 0} \frac{(ka)^p}{p!} \, \partial_x^p \Phi(x)
\ee
and
\be
\label{p-der-field-replacement}
 \hat{p}_{i+k} 
\; \longrightarrow \;
 a\,\Pi(x+ k a)
=\,
a \sum_{p \geqslant 0} \frac{(ka)^p}{p!} \, \partial_x^p \Pi(x)\,.
\ee
In (\ref{H_M asymmetric}) and (\ref{H_N asymmetric})
we find it convenient to insert the UV cutoff into the sums by writing them as
$ L \sum_{i=1}^L (\dots )  = \frac{(L a)}{a^2}\sum_{i=1}^L (\dots ) a$
because $\sum_{i=1}^L (\dots)a  \longrightarrow  \int_0^\ell  (\dots)  dx$ 
in the continuum limit
and the divergent factor $L$ provides the factor $\ell$.
From (\ref{qp-field-replacement}), (\ref{q-der-field-replacement}) and (\ref{p-der-field-replacement}),
for the operators (\ref{H_M asymmetric}) and (\ref{H_N asymmetric})
it is straightforward to obtain 
$\widehat{H}_M \longrightarrow H_M$ and $\widehat{H}_N  \longrightarrow  H_N$
respectively
in the continuum limit, where
\bea
\label{H_M def}
&& \hspace{-.5cm}
H_M
=
\frac{\ell}{a^2}
\int_0^\ell 
\!\bigg( \mu_0(x) \, \Phi(x)^2
+
2 \sum_{k =1}^{k_{\textrm{\tiny max}}} 
\mu_k(x+ka/2)
\, \Phi(x)\, \Phi(x+ k a)
\bigg)
dx 
\\
\label{H_N def}
&& \hspace{-.5cm}
H_N
=
\ell
\int_0^\ell 
\!\bigg( \nu_0(x) \, \Pi(x)^2
+
2 \sum_{k =1}^{k_{\textrm{\tiny max}}} 
\nu_k(x+ka/2)
\, \Pi(x)\, \Pi(x+ k a)
\bigg)
dx 
\eea
being $k_{\textrm{\tiny max}}$ the number of diagonals to include in the sums occurring in
these expressions.

In our numerical analysis the parameter $k_{\textrm{\tiny max}}$ plays a crucial role
which is discussed below in this subsection and also in Appendix\;\ref{sec_app:details}.
Since $k_{\textrm{\tiny max}} \to \infty$ in the continuum limit 
(see (\ref{q-der-field-replacement}) and (\ref{p-der-field-replacement})),
increasing values of $k_{\textrm{\tiny max}}$ are considered. 
We find it worth remarking that the limit $L \to \infty$ in (\ref{munuk def}) is taken before the limit $k_{\textrm{\tiny max}} \to \infty$.
This implies that we have to consider the regime given by $k_{\textrm{\tiny max}} \ll L$
in our numerical studies, where both $L$ and $k_{\textrm{\tiny max}}$ are finite.

Since $a \to 0$ in the continuum limit, we expand the integrands in (\ref{H_M def}) and (\ref{H_N def})
by keeping only the terms that could provide a non vanishing contribution after the limit. 
For (\ref{H_M def}) we obtain
\bea
\label{Moperator expansion1}
& &\hspace{-2.4cm}
H_M
=
\frac{\ell}{a^2}
\int_0^\ell 
\Bigg\{  
\mathcal{M}_{k_\textrm{\tiny max}}^{(0)}(x) \, \Phi(x)^2
 +
a \sum_{k =1}^{k_{\textrm{\tiny max}}} k
\Big[ \,\mu_k'(x) \, \Phi(x) + 2\,\mu_k(x) \, \Phi(x)' \,\Big] \Phi(x)
\\
& & \hspace{1.7cm}
+ a^2
 \sum_{k =1}^{k_{\textrm{\tiny max}}} k^2
\left[ \, \frac{1}{4} \,\mu_k''(x) \, \Phi(x) + \partial_x \Big( \mu_k (x) \, \Phi'(x) \Big)  \,\right] \Phi(x)
\Bigg\} \,dx
+O(a)
\nonumber
\eea
where we have introduced the  function
\be
\label{M-summation k^0 continuum}
\mathcal{M}_{k_\textrm{\tiny max}}^{(0)}(x)
 \equiv
 \lim_{L \to \infty} \frac{ \mathsf{M}_{k_\textrm{\tiny max}}^{(0)}(i)}{L}
  =
 \mu_{0}(x) + 2 \sum_{k=1}^{k_\textrm{\tiny max}} \mu_k(x)
\ee
defined by combining the diagonals of the symmetric matrix $M$ as follows
\be
\label{M-summation k^0 def}
 \mathsf{M}_{k_\textrm{\tiny max}}^{(0)}(i)
 \equiv
M_{i,i} + 2 \sum_{k=1}^{k_\textrm{\tiny max}} M_{i,i+k}\,.
\ee

While the expansion (\ref{Moperator expansion1}) contains terms that are divergent if the corresponding weight functions are non vanishing, 
it is straightforward to notice that (\ref{H_N def}) is finite as $a\to 0$. 
Indeed, its Taylor expansion reads
\be
\label{Noperator expansion1}
H_N
\,=\,
\ell
\int_0^\ell 
\mathcal{N}_{k_\textrm{\tiny max}}^{(0)}(x) \,  \Pi(x)^2 \,dx + O(a)
\ee
where the function $\mathcal{N}_{k_\textrm{\tiny max}}^{(0)}(x) $ 
is the combination of the functions $\nu_k(x)$ in (\ref{munuk def}) given by
\be
\label{N-summation k^0 continuum}
\mathcal{N}_{k_\textrm{\tiny max}}^{(0)}(x)
 \equiv
 \lim_{L \to \infty} \frac{ \mathsf{N}_{k_\textrm{\tiny max}}^{(0)}(i)}{L}
 =
 \nu_{0}(x) + 2 \sum_{k=1}^{k_\textrm{\tiny max}} \nu_k(x)
\ee
being  $ \mathsf{N}_{k_\textrm{\tiny max}}^{(0)}(i)$ the same combination of the corresponding diagonals of the symmetric matrix $N$, namely
\be
\label{N-summation k^0 def}
 \mathsf{N}_{k_\textrm{\tiny max}}^{(0)}(i)
 \equiv
N_{i,i} + 2 \sum_{k=1}^{k_\textrm{\tiny max}} N_{i,i+k} \,.
\ee

Assuming that the integral and the discrete sums can be exchanged
in the expression (\ref{Moperator expansion1}) for $H_M$,
one notices that the integrand of the $O(1/a)$ term is the total derivative
$\partial_x [\mu_k(x)  \, \Phi(x)^2] $;
hence its integration provides the boundary terms $[\mu_k(x)  \, \Phi(x)^2 ] |^{x=\ell}_{x=0}\,$.
These boundary terms vanish because $\mu_k(0)  = \mu_k(\ell)  =0$
for the interval in the infinite line (see Fig.\,\ref{fig:Muk-Infinite}).
As for the $O(1)$ term in (\ref{Moperator expansion1}),
an integration by parts can be performed for the term
whose integrand is $\Phi(x) \,\partial_x [ \mu_k (x) \, \Phi(x)'] $,
and the resulting boundary terms vanish, again because $\mu_k(0) = \mu_k(\ell) =0$.
%%%
By employing these observations and discarding the $O(a)$ terms,
the expression (\ref{Moperator expansion1}) can be written as follows
\be
\label{Moperator expansion2}
H_M
=
\frac{\ell}{a^2}
\int_0^\ell 
\!\left\{  
\mathcal{M}_{k_\textrm{\tiny max}}^{(0)}(x) \, \Phi(x)^2
 + a^2
 \sum_{k =1}^{k_{\textrm{\tiny max}}} k^2 \!
\left[ \, \frac{1}{4} \,\mu_k''(x) \, \Phi(x)^2 -\mu_k (x) \, \big(\Phi'(x)\big)^2 \,\right]
\right\} dx\,.
\ee

Considering the integral whose integrand is $\mu_k (x) [\Phi'(x)]^2$ from the $O(1)$ term of this expression,
we find it worth defining
\be
\label{M-summation k^2 continuum}
\mathcal{M}_{k_\textrm{\tiny max}}^{(2)}(x)
 \equiv
 \lim_{L \to \infty} \frac{ \mathsf{M}_{k_\textrm{\tiny max}}^{(2)}(i)}{L}
 \equiv
 \sum_{k =1}^{k_{\textrm{\tiny max}}} k^2 \mu_{k}(x_k)
\ee
where, by using (\ref{munuk def}), we have introduced
the following combination of diagonals of the symmetric matrix $M$ 
\be
\label{M-summation k^2 def}
  \mathsf{M}_{k_\textrm{\tiny max}}^{(2)}(i)
 \equiv
  \sum_{k =1}^{k_{\textrm{\tiny max}}} k^2 M_{i,i+k}\,.
\ee

As for the integral whose integrand is $\mu_k''(x) \, \Phi(x)^2$ in (\ref{Moperator expansion2}),
we approximate the functions $\mu_k''(x)$ through finite differences
because the analytic expressions of the functions $\mu_k(x)$ are not available.
Thus, we have
\be
\label{second der discrete}
a^2 \mu_k''(x) 
=
\mu_k(x+a) -2\, \mu_k(x)+\mu_k(x-a)\,.
\ee
This expression and (\ref{munuk def}) naturally lead us to introduce
\be
\label{M-summation k^2 der2 continuum}
\mathcal{M}_{2,k_\textrm{\tiny max}}^{(2)}(x)
 \equiv
 \lim_{L \to \infty} \frac{ \mathsf{M}_{2,k_\textrm{\tiny max}}^{(2)}(i)}{L}
 \equiv
 \sum_{k =1}^{k_{\textrm{\tiny max}}} k^2 \mu_{2,k}(x_k)
\ee
where the subindex $2$ indicates that these quantities are related to the second derivative of $\mu_k(x)$.
In (\ref{M-summation k^2 der2 continuum})
we have defined the functions $\mu_{2,k}(x_k)$ as follows
\be
\lim_{L \to \infty}
\frac{M_{i+1, i+1+k} - 2M_{i, i+k} +M_{i-1, i-1+k}}{L}
\,\equiv \,\mu_{2,k}(x_k)
\ee
and the combinations of matrix elements of $M$ given by 
\be
\label{M-summation k^2 der2 def}
\mathsf{M}_{2,k_\textrm{\tiny max}}^{(2)}(i) 
\equiv
 \sum_{k =1}^{k_{\textrm{\tiny max}}} k^2 
 \big(
 M_{i+1, i+1+k} - 2M_{i, i+k} +M_{i-1, i-1+k}
 \big)\,.
\ee

In the continuum limit $k_\textrm{\tiny max} \to \infty$;
hence we introduce the weight functions obtained 
by taking this limit in (\ref{M-summation k^0 continuum}), 
(\ref{N-summation k^0 continuum}),
(\ref{M-summation k^2 continuum}) 
and (\ref{M-summation k^2 der2 continuum}),
namely 
\be
\label{MN-summation k^0 continuum infty}
\mathcal{M}_{k_\textrm{\tiny max}}^{(0)}(x) \,\longrightarrow\, \mathcal{M}_{\infty}^{(0)}(x) 
\;\; \qquad \;\;
\mathcal{N}_{k_\textrm{\tiny max}}^{(0)}(x) \,\longrightarrow\, \mathcal{N}_{\infty}^{(0)}(x) 
\ee
and
\be
\label{M-summation k^2 def kmax-infty}
\mathcal{M}_{k_\textrm{\tiny max}}^{(2)}(x) \,\longrightarrow\, \mathcal{M}_{\infty}^{(2)}(x)
\;\;\qquad\;\;
\mathcal{M}_{2,k_\textrm{\tiny max}}^{(2)}(x) \,\longrightarrow\, \mathcal{M}_{2,\infty}^{(2)}(x)\,.
\ee

Summarising, the continuum limit of the entanglement hamiltonian (\ref{KA/2})
obtained from (\ref{H_M asymmetric}) and (\ref{H_N asymmetric})
is found by taking  the limit $k_\textrm{\tiny max} \to \infty$ of half of the 
sum of (\ref{Noperator expansion1}) and (\ref{Moperator expansion2}).
By employing the functions introduced in  
(\ref{MN-summation k^0 continuum infty}) and (\ref{M-summation k^2 def kmax-infty}), 
for the continuum limit of the entanglement hamiltonian (\ref{KA/2}) we find
\bea
\label{MplusNoperator expansion}
& & \hspace{-1cm}
\frac{H_M + H_N}{2}
\,=\,
\frac{\ell}{a^2}
\int_0^\ell 
\frac{1}{2}\left[ \,\mathcal{M}_{\infty}^{(0)}(x) 
+  \frac{1}{4}\, \mathcal{M}_{2,\infty}^{(2)}(x) \right] 
\Phi(x)^2\, dx\,
\\
\rule{0pt}{.8cm}
& & \hspace{1.4cm}
+\, \ell
\int_0^\ell 
\frac{1}{2}\Big[ \, 
\mathcal{N}_{\infty}^{(0)}(x) \,  \Pi(x)^2 - \mathcal{M}_{\infty}^{(2)}(x) \big(\Phi'(x)\big)^2 
\,\Big] dx
+O(a) \,.
\nonumber
\eea

Since analytic results for the functions $\mu_k$ and $\nu_k$ are not available,
we study the weight functions 
$\mathcal{M}_{\infty}^{(0)}(x) $, $ \mathcal{M}_{2,\infty}^{(2)}(x)$, $\mathcal{N}_{\infty}^{(0)}(x)$ and $ \mathcal{M}_{\infty}^{(2)}(x)$
in (\ref{MplusNoperator expansion})
by performing a numerical analysis of the combinations 
of the matrix elements of $M$ and $N$ defining them,
which are given respectively by
(\ref{M-summation k^0 def}), (\ref{M-summation k^2 der2 def}), 
(\ref{N-summation k^0 def})  and (\ref{M-summation k^2 def}).
These combinations depend on the number of sites $L$ in the interval 
and on the parameter $k_{\textrm{\tiny max}}$
labelling the number of diagonals to include in the sums. 
As already remarked above, we study the continuum limit 
by taking $L \to \infty$ first,
in order to guarantee that the functions $\mu_k$ and $\nu_k$ in (\ref{munuk def}) are well defined,
and then $k_{\textrm{\tiny max}} \to \infty$.
This means that we must keep $k_{\textrm{\tiny max}} \ll L$ in our numerical analysis.
If we had analytic expressions for the functions $\mu_k$ and $\nu_k$,
we could check whether they vanish fast enough as $k \to \infty$ 
to find convergence in the infinite sums defining the weight functions in
(\ref{MN-summation k^0 continuum infty}) and (\ref{M-summation k^2 def kmax-infty}),
which occur in (\ref{MplusNoperator expansion}).

 \begin{figure}[t!]
\vspace{.2cm}
\hspace{-1.3cm}
% \begin{center}
\includegraphics[width=1.12\textwidth]{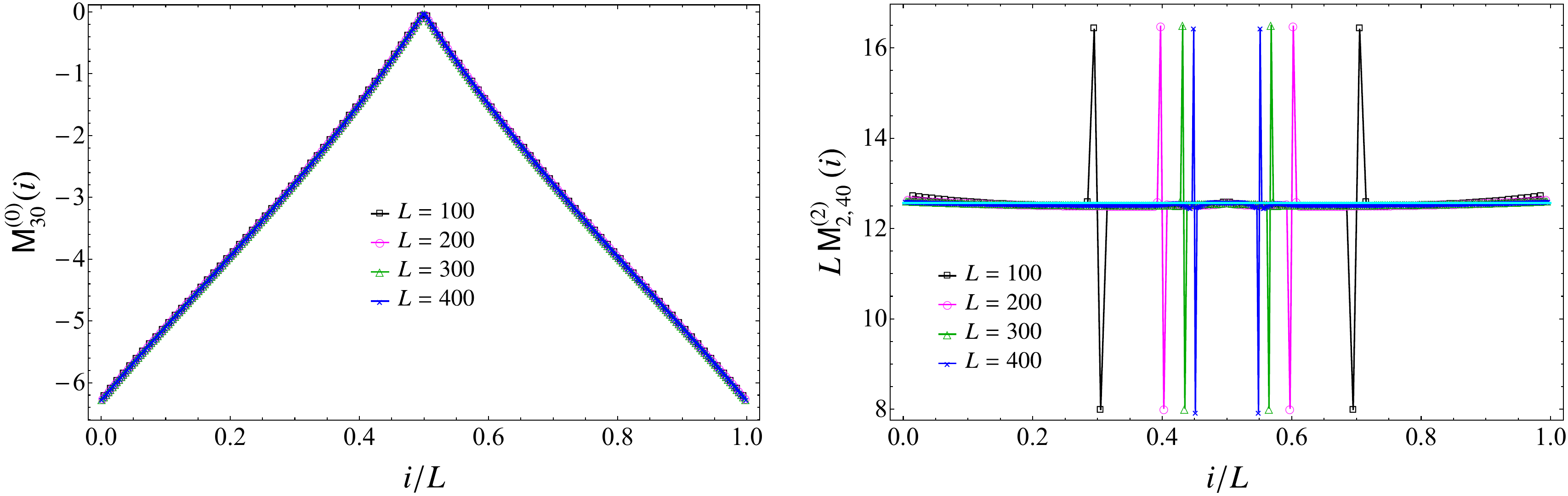}
% \end{center}
\vspace{-.5cm}
\caption{
The combinations (\ref{M-summation k^0 symm}) (left)
and (\ref{M-summation k^2 der2 symm}) (right)
when the subsystem is an interval made by $L$ sites in the infinite line 
and $\omega L =10^{-500}$.
The cyan horizontal line in the right panel corresponds to $4\pi$.
The collapses of the data points for increasing values of $L$ support (\ref{zero-M-functions}).
}
\vspace{.1cm}
\label{fig:Zeros-Infinite}
\end{figure}

It is important to observe that,
since (\ref{M-summation k^0 def}), (\ref{N-summation k^0 def}), (\ref{M-summation k^2 der2 def}) and (\ref{M-summation k^2 def}) 
can be evaluated only in the spatial range given by $1\leqslant i\leqslant L-k_{\textrm{\tiny max}}$,
these combinations are not defined on the whole interval for finite values of $L$ and $k_{\textrm{\tiny max}}$.
%%%%%
The numerical results for these combinations
are shown in the Appendix\;\ref{sec_app:details}
(see the top panels in Fig.\,\ref{fig:sum-methods-M-Infinite} 
and Fig.\,\ref{fig:sum-methods-N-Infinite}):
they do not provide symmetric curves with respect to the center of the interval,
as expected from the symmetry of the configuration,
and they do not capture 
the CFT curve close to the right endpoint of the interval.
%%%
This motivates us to employ decompositions of the operators $\widehat{H}_M$ and $\widehat{H}_N$
that are more suitable than (\ref{H_M asymmetric}) and (\ref{H_N asymmetric})
to obtain the CFT predictions on the entire interval.
The decompositions (\ref{H_M Casini}) and (\ref{H_N Casini}) 
provide curves that are symmetric with respect to the center of the interval, 
but they do not allow to recover the CFT curve close to both the endpoints of the interval
(see the middle panels in Fig.\,\ref{fig:sum-methods-M-Infinite} 
and Fig.\,\ref{fig:sum-methods-N-Infinite}).
In the following we consider the decompositions (\ref{H_M symm}) and (\ref{H_N symm}).

The procedure explained above 
to study the continuum limit of the entanglement hamiltonian
can be adapted straightforwardly to the case where 
the decompositions (\ref{H_M symm}) and (\ref{H_N symm}) are employed.
The result is again (\ref{MplusNoperator expansion}), with the weight functions given by
(\ref{MN-summation k^0 continuum infty}) and (\ref{M-summation k^2 def kmax-infty}).
The crucial difference with respect to the previous analysis is that, as $L \to \infty$,
in (\ref{MN-summation k^0 continuum infty}) we have 
$  \mathsf{M}_{k_\textrm{\tiny max}}^{(0)}(i) / L \rightarrow \mathcal{M}_{k_\textrm{\tiny max}}^{(0)}(x)$ 
with
\bea
\label{M-summation k^0 symm}
\mathsf{M}_{k_\textrm{\tiny max}}^{(0)}(i)
&=&
 \begin{cases}
M_{i,i} + 2 \sum_{k=1}^{k_\textrm{\tiny max}} M_{i,i+k} & \qquad 1\leqslant i\leqslant L/2
\\
\rule{0pt}{.7cm}
M_{i,i} + 2 \sum_{k=1}^{k_\textrm{\tiny max}} M_{i-k,i} & \qquad L/2+1\leqslant i\leqslant L
\end{cases}
\eea
and $  \mathsf{N}_{k_\textrm{\tiny max}}^{(0)}(i) / L \rightarrow \mathcal{N}_{k_\textrm{\tiny max}}^{(0)}(x)$  with
\bea
\label{N-summation k^0 symm}
\mathsf{N}_{k_\textrm{\tiny max}}^{(0)}(i)
&=&
  \begin{cases}
N_{i,i} + 2 \sum_{k=1}^{k_\textrm{\tiny max}} N_{i,i+k} & \qquad 1\leqslant i\leqslant L/2
\\
 \rule{0pt}{.7cm}
N_{i,i} + 2 \sum_{k=1}^{k_\textrm{\tiny max}} N_{i-k,i} & \qquad L/2+1\leqslant i\leqslant L
\end{cases}
\eea
while in (\ref{M-summation k^2 def kmax-infty}) we have
$ \mathsf{M}_{k_\textrm{\tiny max}}^{(2)}(i) / L \to \mathcal{M}_{k_\textrm{\tiny max}}^{(2)}(x)$ 
with
\bea
\label{M-summation k^2 symm}
& & 
 \mathsf{M}_{k_\textrm{\tiny max}}^{(2)}(i)
=
  \begin{cases}
  \sum_{k =1}^{k_{\textrm{\tiny max}}} k^2 M_{i,i+k} & \qquad 1\leqslant i\leqslant L/2
  \\
   \rule{0pt}{.7cm}
  \sum_{k =1}^{k_{\textrm{\tiny max}}} k^2 M_{i-k,i} & \qquad L/2+1\leqslant i\leqslant L
  \end{cases}
\eea
and $ \mathsf{M}_{2,k_\textrm{\tiny max}}^{(2)}(i) / L \to \mathcal{M}_{2,k_\textrm{\tiny max}}^{(2)}(x) $ with
\bea
\label{M-summation k^2 der2 symm}
& & \hspace{-2.2cm}
\mathsf{M}_{2,k_\textrm{\tiny max}}^{(2)}(i) 
=
 \begin{cases}
 \sum_{k =1}^{k_{\textrm{\tiny max}}} k^2 
 \big(
 M_{i+1, i+1+k} - 2M_{i, i+k} +M_{i-1, i-1+k} \big) & \quad 1\leqslant i\leqslant L/2
 \\
 \rule{0pt}{.7cm}
 \sum_{k =1}^{k_{\textrm{\tiny max}}} k^2 
 \big(
 M_{i-k+1, i+1} - 2M_{i-k, i} +M_{i-k-1, i-1}
 \big) & \quad L/2+1\leqslant i\leqslant L\,.
 \end{cases}
\eea
The occurrence of two branches in these functions of the spatial index $i$ 
(which originates from the splitting of the range $1\leqslant i \leqslant L$ in (\ref{H_M symm}) and (\ref{H_N symm}))
guarantees that they are well defined on the entire interval for finite values of $k_{\textrm{\tiny max}} \ll L$.
The combinations of diagonals in 
(\ref{M-summation k^0 symm}), (\ref{N-summation k^0 symm}), 
(\ref{M-summation k^2 symm}) and (\ref{M-summation k^2 der2 symm})
display the symmetry under reflection with respect to the center of the interval, 
which has been observed also for the diagonals of $M$ and $N$
(see Fig.\,\ref{fig:Muk-Infinite} and Fig.\,\ref{fig:Nuk-Infinite}).

\begin{figure}[t!]
\vspace{.2cm}
\hspace{-1.3cm}
% \begin{center}
\includegraphics[width=1.12\textwidth]{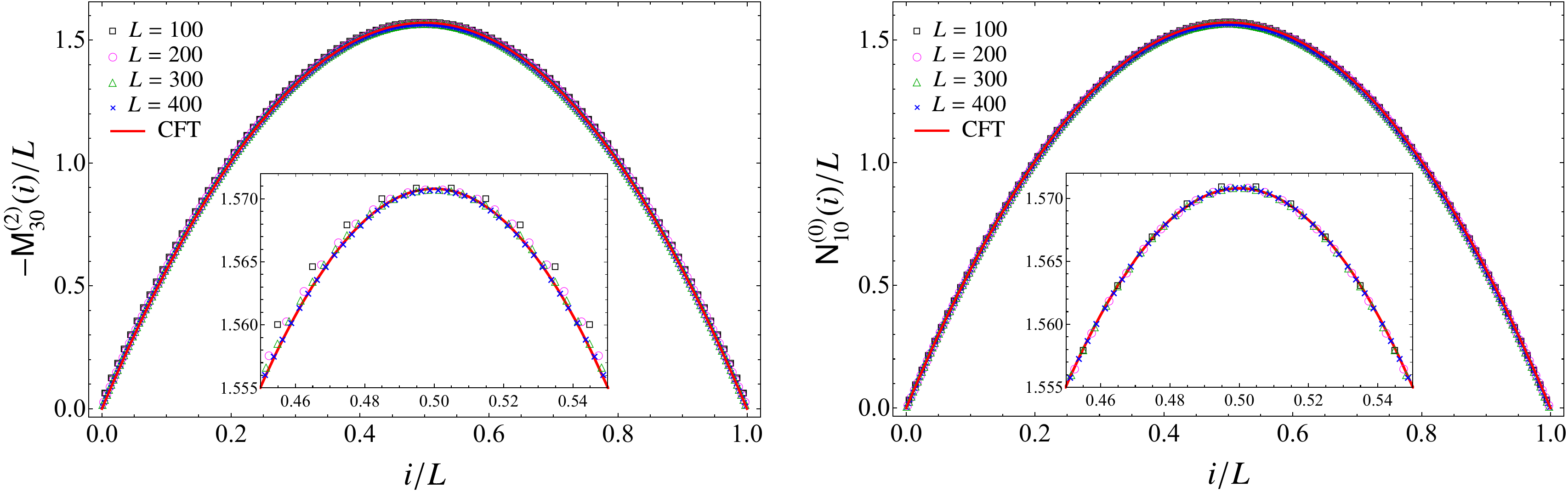}
% \end{center}
\vspace{-.5cm}
\caption{
The combinations (\ref{M-summation k^2 symm}) (left)
and (\ref{N-summation k^0 symm}) (right)
when the subsystem is an interval made by $L$ sites in the infinite line 
and $\omega L =10^{-500}$.
The collapses of the data points corresponding to increasing values of $L$ support
(\ref{MN-beta-functions}),
with $\beta(x)$ given by the parabola (\ref{parabola-CFT-infinite}) (red solid curve).
}
\vspace{.0cm}
\label{fig:Parabola-Infinite}
\end{figure}

In Fig.\,\ref{fig:Zeros-Infinite} we show some numerical results 
for the combinations in (\ref{M-summation k^0 symm}) and (\ref{M-summation k^2 der2 symm}).
From the left panel we observe that, when $k_{\textrm{\tiny max}}$ is large enough,
$\mathsf{M}_{k_\textrm{\tiny max}}^{(0)}$ converges to a well defined function of $x/\ell \in (0,1)$.
This observation allows to conclude that $\mathsf{M}_{k_\textrm{\tiny max}}^{(0)} / L \to 0$ as $L \to \infty$
at any given value of $x/\ell \in (0,1)$.
Similarly, the data reported in the right panel of Fig.\,\ref{fig:Zeros-Infinite} show that, 
when $k_{\textrm{\tiny max}}$ is large enough,
the product $L \,\mathsf{M}_{2,k_\textrm{\tiny max}}^{(2)}$ for increasing values of $L$ 
collapses on the horizontal line corresponding to $4\pi$
except for four isolated and finite picks in each curve, 
whose positions depend on $L$ and whose heights are independent of $L$. 
The positions of these picks are symmetric with respect to the center of the interval
and they move towards the center of the interval as $L$ increases.
These observations allow to conclude that $\mathsf{M}_{2,k_\textrm{\tiny max}}^{(2)} / L \to 0$ as $L \to \infty$.  
Thus, the collapses of the data points observed in Fig.\,\ref{fig:Zeros-Infinite} for increasing values of  $L$
lead to conclude that for the weight functions occurring in the $O(1/a^2)$ term of (\ref{MplusNoperator expansion}) we should have
\be
\label{zero-M-functions}
\mathcal{M}_{\infty}^{(0)}(x) = 0
\;\;\qquad \;\;
\mathcal{M}_{2,\infty}^{(2)}(x) = 0\,.
\ee
The curves in Fig.\,\ref{fig:Zeros-Infinite} are obtained through the decompositions (\ref{H_M symm}) and (\ref{H_N symm}).
Considering the  other decompositions reported in \S\ref{sec:EHinHC} and in Appendix\;\ref{sec_app:details},
one finds different curves, but all of them lead to the CFT prediction (\ref{zero-M-functions}).

 \begin{figure}[t!]
\vspace{.2cm}
\hspace{-1.3cm}
% \begin{center}
\includegraphics[width=1.12\textwidth]{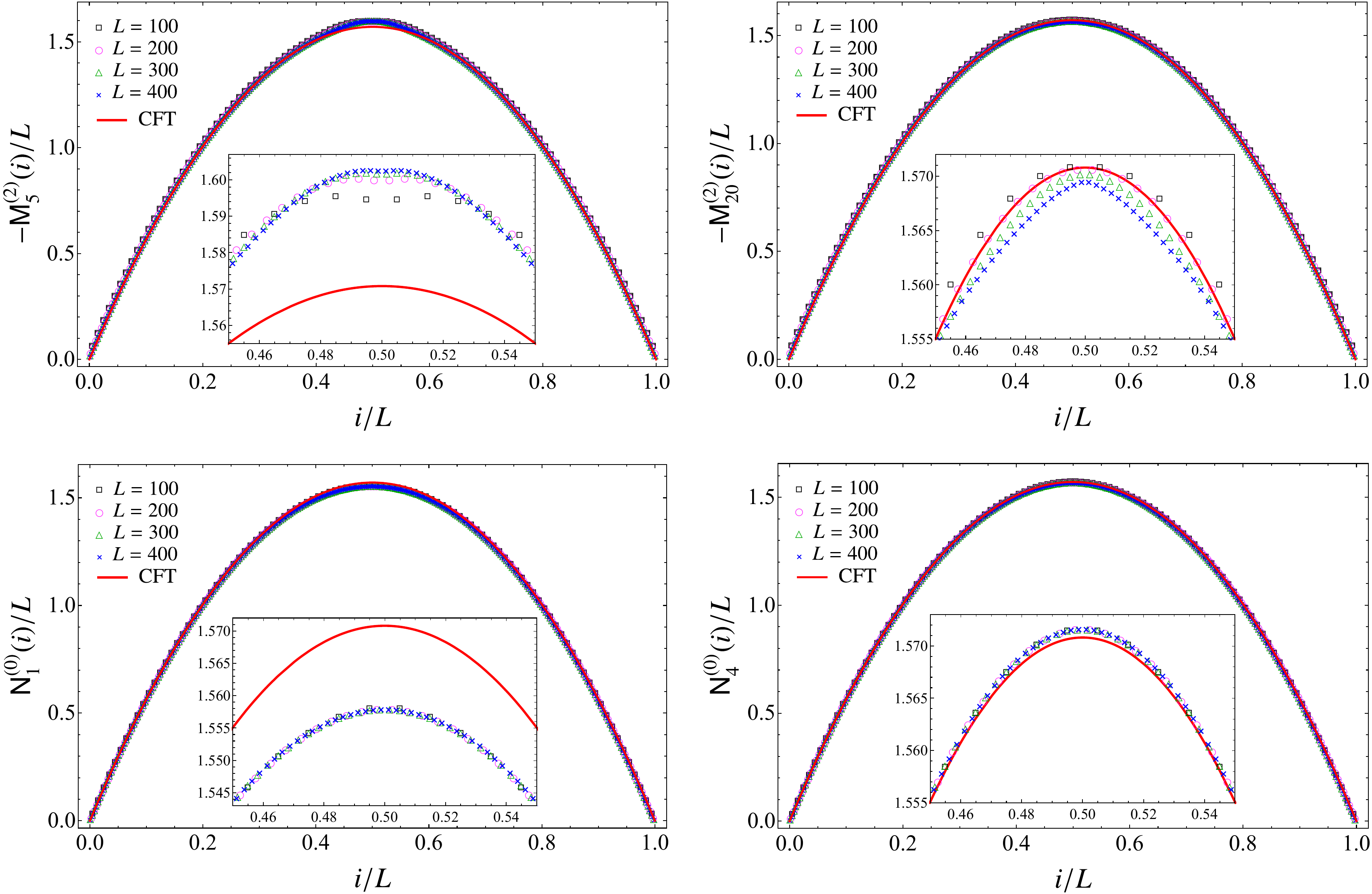}
% \end{center}
\vspace{-.5cm}
\caption{
Role of the parameter $k_{\textrm{\tiny max}}$
in the combinations (\ref{M-summation k^2 symm}) (top panels)
and (\ref{N-summation k^0 symm}) (bottom panels)
when the subsystem is an interval made by $L$ sites in the infinite line 
and $\omega L =10^{-500}$.
The insets, which zoom in on the central part of the interval,
show that the agreement with the CFT prediction
given by the parabola (\ref{parabola-CFT-infinite}) (red solid curve)
improves as $k_{\textrm{\tiny max}}$ increases.
}
\vspace{.0cm}
\label{fig:kmax-Symm-Infinite}
\end{figure}

In Fig.\,\ref{fig:Parabola-Infinite} and Fig.\,\ref{fig:kmax-Symm-Infinite} 
we report  numerical results for the combinations in (\ref{N-summation k^0 symm}) and (\ref{M-summation k^2 symm}).
\\
Comparing these two figures, it is straightforward to conclude 
that the agreement between the numerical data and the CFT prediction $\beta(x)$
given by the parabola (\ref{parabola-CFT-infinite}) (red solid curve)
improves as $k_{\textrm{\tiny max}} \ll L$ increases,
i.e. by including more diagonals in the sums occurring in 
(\ref{N-summation k^0 symm}) and (\ref{M-summation k^2 symm}).
The data reported  in  Fig.\,\ref{fig:Zeros-Infinite} and Fig.\,\ref{fig:Parabola-Infinite}
correspond to the optimal values of $k_{\textrm{\tiny max}}$,
when the behaviours of the data become stable.
These optimal values are different for the combinations involving $M$ and $N$.
From Fig.\,\ref{fig:kmax-Symm-Infinite} 
we also observe a parity effect in $k$:
the asymptotic curve for a given $k_{\textrm{\tiny max}}$ 
is either above or below the CFT curve, 
depending on the parity of $k_{\textrm{\tiny max}}$, 
and the distances between these curve decrease as $k_{\textrm{\tiny max}}$ increases
until the optimal value is reached. 
We remark that the data points reported in 
Fig.\,\ref{fig:Zeros-Infinite}, Fig.\,\ref{fig:Parabola-Infinite} and Fig.\,\ref{fig:kmax-Symm-Infinite}
probe the entire interval $A$, 
including the neighbourhoods of the endpoints.
Furthermore, the resulting curves are symmetric under 
reflection with respect to the center of the interval, 
as expected for this bipartition.  
These features support our choice to employ the combinations
(\ref{M-summation k^0 symm}), (\ref{N-summation k^0 symm}), 
(\ref{M-summation k^2 symm}) and (\ref{M-summation k^2 der2 symm}).

The collapses of the data points in Fig.\,\ref{fig:Parabola-Infinite} 
for increasing values of $L$ lead to conjecture 
that the weight functions occurring in the finite term of (\ref{MplusNoperator expansion})
are
\be
\label{MN-beta-functions}
 \mathcal{M}_{\infty}^{(2)}(x) 
 = -\, \beta(x)
\hspace{.6cm} \qquad \hspace{.6cm}
 \mathcal{N}_{\infty}^{(0)}(x) 
 =  \beta(x)\,.
\ee

Thus, by employing the numerical results 
(\ref{zero-M-functions}) and (\ref{MN-beta-functions})
into the expression (\ref{MplusNoperator expansion})
for the entanglement hamiltonian of an interval in the infinite line,
we find the CFT prediction (\ref{EH-intro}) 
with $\beta(x)$ given by  (\ref{parabola-CFT-infinite}) 
and the energy density by (\ref{T00 massless infinite}).

We find it worth remarking that the height of the cyan horizontal line  
in the right panel of Fig.\,\ref{fig:Zeros-Infinite} 
corresponds to $4\pi = -\,\ell^2 \beta(x)'' $,  being 
$\beta(x)$ the weight function (\ref{parabola-CFT-infinite})
predicted by CFT.
A naive explanation of this observation comes from the fact that 
$\mathsf{M}_{2,k_\textrm{\tiny max}}^{(2)}$ is a combination obtained 
through a finite differences approximation of $\mu_k''(x)$ (see (\ref{M-summation k^2 der2 def}))
and that the combination $\mathsf{M}_{k_\textrm{\tiny max}}^{(2)}$ of $\mu_k(x)$
provides $-\beta(x)$ in the continuum limit (see (\ref{MN-beta-functions})).
Nonetheless, an exchange of the second derivate of $\mu_k(x)$ with the discrete sum over $k$ in (\ref{Moperator expansion1})
would provide an unexpected term containing $\Phi^2$ multiplied by a constant weight function in the entanglement hamiltonian.
This leads us to conclude that exchanges between derivatives with respect to $x$ and discrete sums over $k$ are not allowed.

%%%%%%%%%%%%%%%%%%%%%%%%%%%%%%%%%%%%%%%%%%%%%%%%%%%%

\subsection{Entanglement spectrum}
\label{sec:infinite-line-gaps}

In a two dimensional CFT,
the entanglement spectra of an interval for the bipartitions shown in Fig.\,\ref{fig:config}
have been studied in \cite{ct-16}  through methods of Boundary Conformal Field Theory (BCFT) 
\cite{bcft, callan-klebanov-94, blumenhagen}.
The occurrence of boundaries is due to the regularisation procedure, as briefly mentioned in \S\ref{sec:intro}.
In the imaginary time description of the two dimensional spacetime underlying the bipartitions shown in Fig.\,\ref{fig:config},
the UV cutoff $\epsilon$ can be introduced 
by removing an infinitesimal disk of radius $\epsilon$ around each entangling point
\cite{wil-larsen-94, peschel-04-eh-fermions, tachikawa, ct-16}. 
For the interval in the infinite line (left panel of Fig.\,\ref{fig:config}),
the remaining spacetime has two boundaries which encircle the two endpoints of the interval $A$;
hence it can be mapped into an annulus through a conformal transformation. 
Given the symmetry of this bipartition with respect to the center of the interval, 
the same conformal boundary condition must be imposed on the two boundaries.

\begin{figure}[t!]
\vspace{.2cm}
\hspace{-1.3cm}
% \begin{center}
\includegraphics[width=1.12\textwidth]{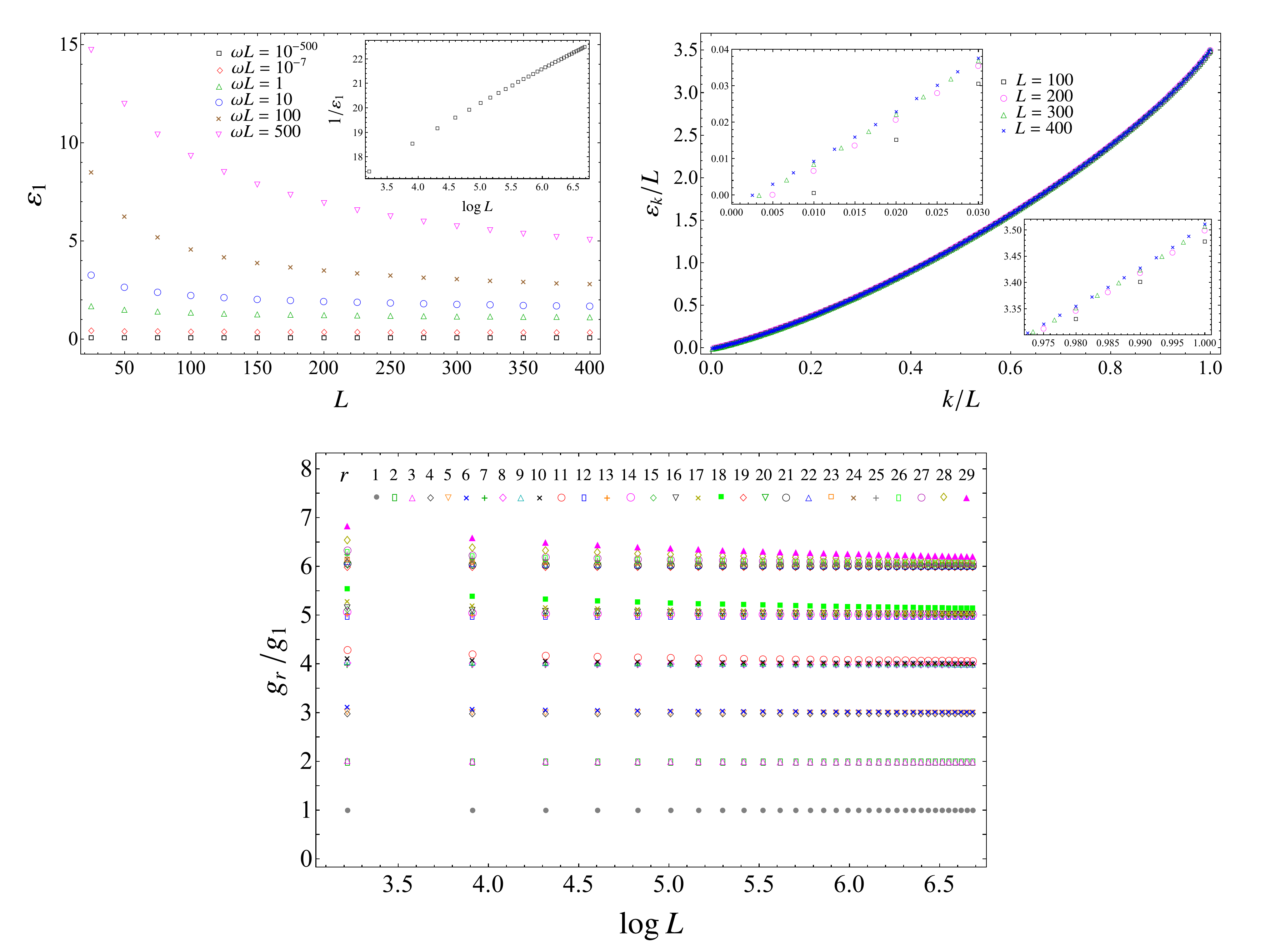}
% \end{center}
\vspace{-.5cm}
\caption{
Entanglement spectrum for an interval in the infinite line.
Top left: The smallest single particle entanglement energy $\varepsilon_1$ 
as function of the number of sites $L$ of the interval, for increasing values of $\omega L$.
Top right: The single particle entanglement energies $\varepsilon_k$ for different values of $L$
(the insets zoom in on the lowest and on the highest values of $k$).
Bottom: The ratios of the gaps $g_r$ in the entanglement spectrum as functions of $\log L$
in the massless regime, i.e. when $\omega L = 10^{-500}$.
}
\vspace{.0cm}
\label{fig:gaps-Infinite}
\end{figure}

For harmonic chains in Gaussian states,
standard techniques allow to evaluate the entanglement spectrum
in terms of the single particle entanglement energies $\varepsilon_r$,
which are obtained from the symplectic spectrum 
of the reduced covariance matrix of the subsystem
\cite{ep-rev, ch-rev, Peschel-eh-free-models, dat-19, sierra-cirac-18}.
%%%
Once the single particle entanglement energies have been ordered 
as $\varepsilon_1 \leqslant \varepsilon_2 \leqslant \dots \leqslant \varepsilon_L$,
the gaps $g_r$ introduced in \S\ref{sec:intro} can be written as linear combinations 
$\sum_{k=1}^L n_k \, \varepsilon_k$ with non negative integer coefficients $n_k$.

In the top left panel of Fig.\,\ref{fig:gaps-Infinite}, we report some numerical results for $\varepsilon_1$.
For a given finite value of $L$, we observe that $\varepsilon_1 \to 0$ as  $\omega L \to 0$,
while this does not happen for $\varepsilon_r$ with $r>1$.
This leads us to assume that $\varepsilon_1$ vanishes 
in the comparison of the numerical data with the CFT predictions in the bottom panel of Fig.\,\ref{fig:gaps-Infinite}. 
In the top right panel of Fig.\,\ref{fig:gaps-Infinite}
we show the single particle entanglement energies 
$\varepsilon_k/L$ in terms of $k/L$ for some values of $L$
and, if $L$ is large enough, we find that
the data having different $L$ collapse on a well defined curve,
that would be interesting to obtain analytically. 
Given the above assumption about $\varepsilon_1$,
in the bottom panel of Fig.\,\ref{fig:gaps-Infinite} we show the numerical data
for the ratios $g_r / g_1$ of the gaps with respect to the first gap as functions of $\log L$.
%for $1 \leqslant r \leqslant 29$.
It is remarkable to observe that, as $L$ increases, 
the values of $g_r / g_1$ with $1 \leqslant r \leqslant 29$
collapse on all the integers $n$ with $1\leqslant n \leqslant 6$
(we checked that $g_r / g_1 > 6.5$ when $r \geqslant 30$ for the largest value of $L$ at our disposal).
This originates from the fact that the single particle entanglement energies
 in the low-lying part of the spectrum are equally separated by a multiple integer of $\varepsilon_2$.
Furthermore, the degeneracy 
of the $n$-th level is given by the number of possible ways to partition the integer $n$.

These numerical results for $g_r / g_1$ are compatible with the conformal spectrum of 
the BCFT given by a free massless scalar field on the segment 
with either Dirichlet or Neumann boundary conditions imposed on both the endpoints
of the segment \cite{blumenhagen},
but they cannot discriminate between these two possibilities. 
This ambiguity can be resolved by considering the 
entanglement spectrum of an interval at the beginning of a semi-infinite line
with Dirichlet boundary conditions, that will be discussed in \S\ref{sec:semi-infinite-line-gaps}.
In terms of the primary fields and of their descendants, 
we observe the towers of the identity and of $\partial \Phi$.

The agreement found above with the spectrum of the BCFT of the 
free massless scalar field is expected only for the low-lying part of the entanglement spectrum.

%\newpage
%%%%%%%%%%%%%%%%%%%%%%%%%%%%%%%%%%%%%%%%%%%%%%%%%%%%%%%%
%%%%%%%%%%%%%%%%%%%%%%%%%%%%%%%%%%%%%%%%%%%%%%%%%%%%%%%%
\section{Interval at the beginning of the semi-infinite line with Dirichlet b.c.}
\label{sec:semi-infinite-line}

In this section we study the continuum limit of the entanglement hamiltonian 
of $L$ consecutive sites at the beginning of 
the massless harmonic chain on the semi-infinite line
with Dirichlet boundary conditions at its endpoint. 
In \S\ref{sec:semi-infinite-line-correlators} we find
analytic expressions for the two-point correlators 
at a generic value of the mass parameter.
Focussing on the massless regime, in \S\ref{sec:semi-infinite-line-massless} 
we adapt the procedure explained in \S\ref{sec:infinite-line} to this case,
finding the CFT prediction (\ref{EH-intro}),
with the weight function (\ref{parabola-CFT-semi-infinite}) 
and the energy density (\ref{T00 massless semi-infinite}).
The continuum limit of the entanglement spectrum is discussed in \S\ref{sec:semi-infinite-line-gaps}.

\subsection{Correlators}
\label{sec:semi-infinite-line-correlators}

A finite harmonic chain in a segment with Dirichlet boundary conditions imposed at the endpoints 
is defined by (\ref{HC ham}) and 
by $\hat{q}_0 = \hat{q}_{\mathcal{L}} = \hat{p}_0 = \hat{p}_{\mathcal{L}} = 0$.
The two-point correlators $ \langle \hat{q}_i \hat{q}_j  \rangle $ and $\langle \hat{p}_i \hat{p}_j  \rangle$ in the ground state 
read respectively \cite{lievens-pbc-08}
\bea
\label{qq open}
& &
\langle \hat{q}_i \hat{q}_j  \rangle =
\frac{1}{\mathcal{L}} \sum_{k=1}^{\mathcal{L}-1} \frac{1}{m \tilde{\omega}_k} \, 
\sin(\pi k\, i/\mathcal{L})  \, \sin(\pi k\, j/\mathcal{L}) 
\\
\rule{0pt}{.8cm}
\label{pp open}
& &
\langle \hat{p}_i \hat{p}_j  \rangle =
\frac{1}{\mathcal{L}} \sum_{k=1}^{\mathcal{L}-1} m \tilde{\omega}_k \, 
\sin(\pi k \, i/\mathcal{L})  \, \sin(\pi k \, j/\mathcal{L}) 
\eea
where the dispersion relation is
\be
\tilde{\omega}_k \equiv 
\sqrt{\omega^2 +\frac{4\kappa}{m}\, \big[ \sin(\pi k/(2\mathcal{L})) \big]^2} \,>\,\omega
\qquad
1 \leqslant k \leqslant \mathcal{L}-1\,.
\ee

In contrast with the harmonic chain in the infinite line (see \S\ref{sec:infinite-line-corr}),
this harmonic chain is not translation invariant;
hence the zero mode does not occur and 
the massless limit $\omega \to 0$ is well defined because
the correlators (\ref{qq open}) and (\ref{pp open}) are finite.

In the thermodynamic limit, the correlators  (\ref{qq open}) and (\ref{pp open}) 
can be written respectively as follows
\bea
\label{qq therm open int}
\langle \hat{q}_i \hat{q}_j \rangle
&=&
\frac{1}{\pi\, m} \int_0^\pi \!\!
\frac{\sin(\theta\, i) \, \sin(\theta\, j)}{\sqrt{\omega^2+ (4\kappa/m)  \big[ \sin(\theta/2)\big]^2}} \;  d\theta
\\
\rule{0pt}{.9cm}
\label{pp therm open int}
\langle \hat{p}_i \hat{p}_j \rangle
&=&
\frac{m}{\pi} \int_0^\pi \!
 \sqrt{\omega^2+ \frac{4\kappa}{m} \, \big[ \sin(\theta/2)\big]^2}
\; \sin(\theta\, i) \, \sin(\theta\, j)\, d\theta
\eea
where $i, j \geqslant 0$ and the Dirichlet boundary conditions are satisfied at the beginning of the semi-infinite line. 
We can evaluate the integrals (\ref{qq therm open int}) and (\ref{pp therm open int}) 
analytically by employing a prosthaphaeresis formula 
and an integral representation of the hypergeometric function\footnote{The following 
integral representation for the hypergeometric function has been employed
\be
\label{HypergeomIntReps}
\int_0^\pi \!\!
\frac{ \cos(n \theta)}{2\pi(1-a \cos\theta)^b} \,  d\theta
\,=\,
 \frac{2^{b-1}\, \Gamma(n+b)}{n!\, a^b\, \Gamma(b)} 
 \left( \frac{1-\sqrt{1-a^2}}{a}\,\right)^{n+b}
  \!\! \, _2F_1 \!\left( b\, ,n+b\, ; n+1 \,; \bigg( \frac{1-\sqrt{1-a^2}}{a}\,\bigg)^2 \,\right)
  \nonumber
\ee
in the special case given by $a=2\zeta /(1+\zeta^2)$ and $b=\pm 1 /2$.
}.
The final result reads
\bea
\label{qq therm open}
\langle \hat{q}_i \hat{q}_j \rangle
&=&
\frac{1}{m\omega \,\sqrt{1+\kappa_\omega}} \;
\Big\{
F_{+}(|i-j|) - F_{+}( i+j)
\Big\}
\\
\label{pp therm open}
\rule{0pt}{.7cm}
\langle \hat{p}_i \hat{p}_j \rangle
&=&
m\omega \,\sqrt{1+\kappa_\omega} \;
\Big\{
F_{-}(|i-j|) - F_{-}( i+j)
\Big\}
\eea
where the functions $F_{\pm}(n)$ are defined as follows
\be
F_{\pm}(n) 
\equiv
\left( \frac{2(1+\kappa_\omega)}{\kappa_\omega}\right)^{\pm 1/2} 
\frac{\Gamma(n\pm 1/2)\, \zeta^{n\pm 1/2}}{n! \, 2\,\Gamma(\pm 1/2)}
\;_2F_1 \big( \pm 1/2\, , n \pm 1/2\, , n+1 \,,  \zeta^2 \,\big)
\ee
with $\kappa_\omega \equiv 2\kappa / (m\omega^2)$ and $\zeta$ given by (\ref{z-def}).

In the massless regime, which corresponds to $\omega = 0$, 
the expressions (\ref{qq therm open}) and (\ref{pp therm open}) 
significantly simplify and become
the correlators found in \cite{cct-neg-long},
which are written in terms of the digamma function $\psi(z)$ respectively as 
\bea
\label{corr qq dirichlet thermo}
& &
\langle \hat{q}_i \hat{q}_j  \rangle =
\frac{1}{2\pi \, \sqrt{\kappa m}} 
\Big(  \psi(1/2+i+j) -  \psi(1/2+i-j)  \Big)
\\
\label{corr pp dirichlet thermo}
\rule{0pt}{.9cm}
& &
\langle \hat{p}_i \hat{p}_j  \rangle =
\frac{2\,\sqrt{\kappa m}}{\pi}  \left( \frac{1}{4(i+j)^2-1} - \frac{1}{4(i-j)^2-1} \right) .
\eea

By restricting the indices $i$ and $j$ of the correlators (\ref{corr qq dirichlet thermo}) 
and (\ref{corr pp dirichlet thermo}) to the interval $A$
at the beginning of the semi-infinite line (see the right panel of Fig.\,\ref{fig:config}), 
i.e. to the integer values in $[1, L]$,
we get the reduced correlation matrices $Q_A$ and $P_A$
to employ in the expression (\ref{eh-block-ch-version}) for
the entanglement hamiltonian matrix $H_A$.
Plugging this matrix into (\ref{ent-ham HC}), we can obtain 
the entanglement hamiltonian of the interval $A$ at the beginning of the 
semi-infinite line with Dirichlet boundary conditions imposed at its origin.

%%%%%%%%%%%%%%%%%%%%%%%%%%%%%%%%%%%%%%%%%%%%%%%%%%%%
\subsection{Entanglement hamiltonian}
\label{sec:semi-infinite-line-massless}

We are interested in the bipartition of the semi-infinite line whose origin coincides with the 
left endpoint of the interval $A$ made by $L$ sites
(see the right panel of Fig.\,\ref{fig:config}), when the entire system is in its ground state. 
The entanglement entropy of this bipartition has been studied for various systems e.g. in 
\cite{cc-04, sierra-taddia}.
In the massless harmonic chain with Dirichlet boundary conditions,
the entanglement hamiltonian $\widehat{K}_A = (\widehat{H}_M + \widehat{H}_N)/2$ 
of this interval is given by 
(\ref{ent-ham HC}) and (\ref{eh-block-ch-version}), 
where the $L \times L$ matrices $Q_A$ and $P_A$ are the reduced correlation matrices 
introduced in \S\ref{sec:semi-infinite-line-correlators}
through the correlators (\ref{corr qq dirichlet thermo}) and (\ref{corr pp dirichlet thermo}).
The result for the continuum limit predicted by CFT is (\ref{EH-intro}),
with the weight function $\beta(x)$ given by (\ref{parabola-CFT-semi-infinite})
and the energy density (\ref{T00 massless semi-infinite}) 
\cite{mintchev-liguori}.
In the following we discuss a numerical procedure to obtain this CFT result.

\begin{figure}[t!]
\vspace{.2cm}
\hspace{-.9cm}
% \begin{center}
\includegraphics[width=1.05\textwidth]{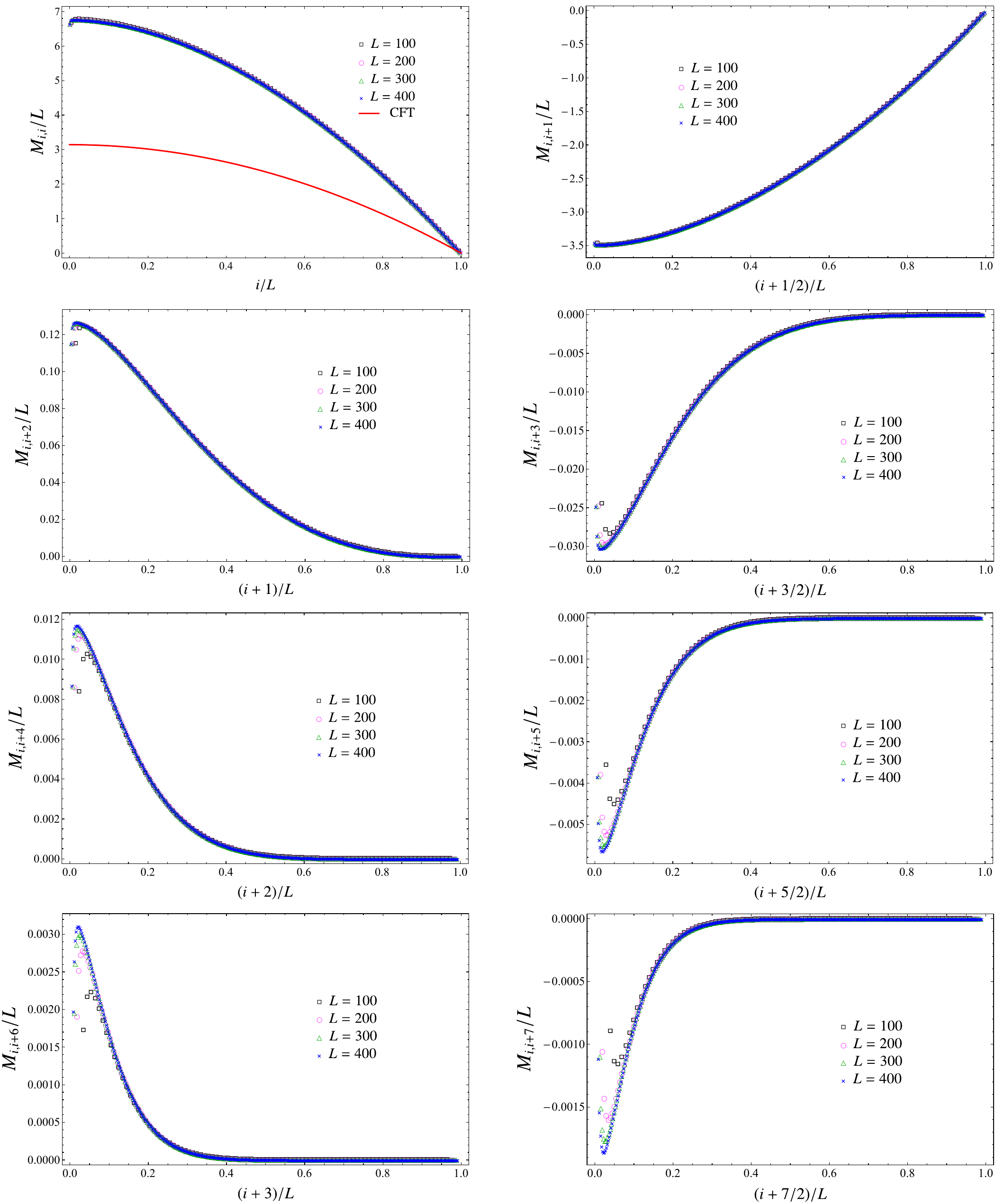}
% \end{center}
\vspace{-.3cm}
\caption{
Diagonals of the  matrix $M$ (see (\ref{munuk def}))
when $A$ is an interval with $L$ sites 
at the beginning of  the semi-infinite line
and $\omega =0$.
The red solid curve is the half parabola (\ref{parabola-CFT-semi-infinite}).
}
\vspace{.1cm}
\label{fig:Muk-Semi-Infinite}
\end{figure}

\begin{figure}[t!]
\vspace{.2cm}
\hspace{-.9cm}
% \begin{center}
\includegraphics[width=1.05\textwidth]{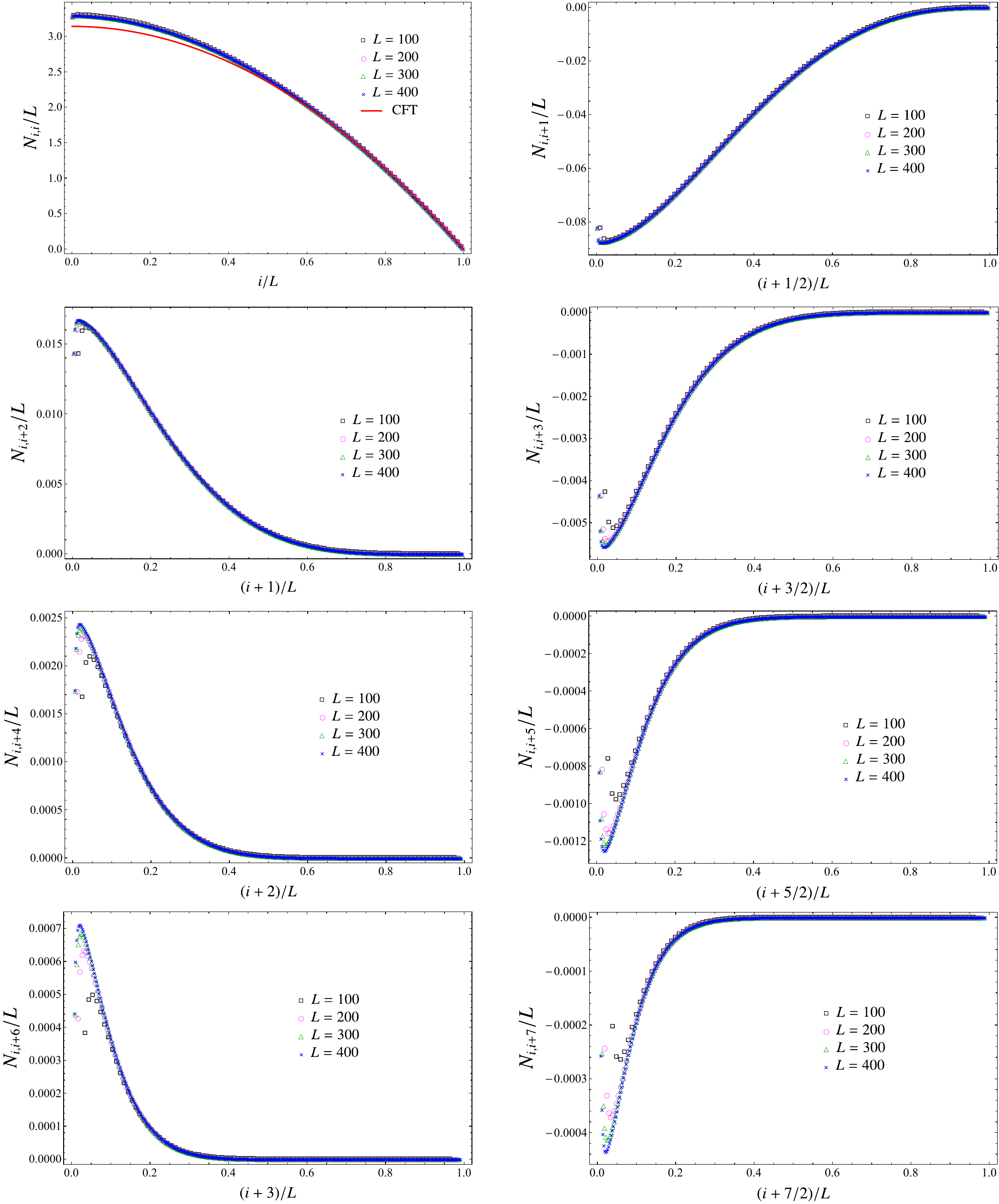}
% \end{center}
\vspace{-.3cm}
\caption{
Diagonals of the  matrix $N$ (see (\ref{munuk def}))
when $A$ is an interval with $L$ sites 
at the beginning of  the semi-infinite line
and $\omega =0$.
The red solid curve is the half parabola (\ref{parabola-CFT-semi-infinite}).
}
\vspace{.4cm}
\label{fig:Nuk-Semi-Infinite}
\end{figure}

The decompositions of the operators $\widehat{H}_M$ and $\widehat{H}_N$
introduced in \S\ref{sec:EHinHC} and in Appendix\;\ref{sec_app:details}
naturally lead to consider the $k$-th diagonals of the symmetric matrices $M$ and $N$, 
like in the case of the interval in the infinite line discussed in \S\ref{sec:infinite-line-massless}.
In the massless regime, we find numerical evidence that the limiting procedures defined in (\ref{munuk def}) 
provide well defined functions for any given value of $k$.
This is shown in Fig.\,\ref{fig:Muk-Semi-Infinite} and Fig.\,\ref{fig:Nuk-Semi-Infinite} 
for the $k$-th diagonal of the matrices $M$ and $N$ respectively, with $0 \leqslant k \leqslant 7$.
Notice that, while the functions $\mu_k$ and $\nu_k$ 
vanish at the entangling point that separates $A$ and $B$,
they are non vanishing at the beginning of the semi-infinite line,
where the Dirichlet boundary condition is imposed. 
It would be interesting to find analytic expressions for these functions. 
Like for the interval in the infinite line, 
these functions have a well defined sign given by the parity of $k$
and the absolute value of their maximum significantly decreases as $k$ increases.

Assuming the existence of the functions $\mu_k$ and $\nu_k$ defined in (\ref{munuk def}),
the continuum limit of the entanglement hamiltonian (\ref{KA/2})
can be studied by adapting to the bipartition that we are considering
the procedure described in \S\ref{sec:infinite-line}.
Special care must be devoted to the boundary terms due to the integrations of a total derivative or 
to the integrations by parts.
In particular, in (\ref{Moperator expansion1})
the integrand of the $O(1/a)$ term is the total derivative $\partial_x [\mu_k(x)  \, \Phi(x)^2] $,
whose integral over the interval gives the boundary terms $[\mu_k(x)  \, \Phi(x)^2 ] |^{x=\ell}_{x=0}\,$.
These boundary terms vanish because $ \mu_k(\ell)  =0$ at the entangling point
and the Dirichlet boundary condition $\Phi(0) = 0$ is imposed at the beginning of the semi-infinite line,
where $\mu_k(0) \neq 0$.
The remaining expression reads
\bea
\label{Moperator expansion1 bdy}
& &\hspace{-.9cm}
H_M
=
\frac{\ell}{a^2}
\int_0^\ell 
\mathcal{M}_{k_\textrm{\tiny max}}^{(0)}(x) \, \Phi(x)^2\, dx
\\
& & \hspace{.3cm}
+ \,\ell
\int_0^\ell \;
 \sum_{k =1}^{k_{\textrm{\tiny max}}} k^2
\left[ \, \frac{1}{4} \,\mu_k''(x) \, \Phi(x) 
+  \mu_k' (x) \, \Phi'(x)  + \mu_k (x) \, \Phi''(x)   \,\right] \Phi(x) \,dx
\nonumber
\eea
where $O(a)$ terms have been discarded 
and $\mathcal{M}_{k_\textrm{\tiny max}}^{(0)}(x)$ has been introduced in (\ref{M-summation k^0 continuum}).

The $O(1)$ term in (\ref{Moperator expansion1 bdy}) is similar to the $O(1)$ term in (\ref{Moperator expansion1})
and the terms containing $\mu_k''(x)$ and $\mu_k (x) $ can be treated as discussed in \S\ref{sec:infinite-line-massless}.
As for the term whose integrand is $\mu_k' (x) \, \Phi'(x)\, \Phi(x)$, 
we approximate $\mu_k' (x)$ through finite differences by writing 
$\mu_k'(x) = [\mu_k(x+a) - \mu_k(x)]/a$
because analytic expressions for $\mu_k (x)$ are not known.
Combining this approximation with (\ref{munuk def}) and (\ref{Moperator expansion1 bdy}),
we are naturally led to introduce 
\be
\label{M-summation k^2 der1 def}
\mathsf{M}_{1,k_\textrm{\tiny max}}^{(2)}(i) 
\equiv
 \sum_{k =1}^{k_{\textrm{\tiny max}}} k^2 
 \big(
 M_{i+1, i+1+k} - M_{i, i+k} 
 \big)
\ee
and
\be
\label{M-summation k^2 der1 continuum}
\mathcal{M}_{1,k_\textrm{\tiny max}}^{(2)}(x)
 \equiv
 \lim_{L \to \infty} \frac{ \mathsf{M}_{1,k_\textrm{\tiny max}}^{(2)}(i)}{L}
 \equiv
 \sum_{k =1}^{k_{\textrm{\tiny max}}} k^2 \mu_{1,k}(x_k)
\ee
being 
\be
\lim_{L \to \infty}
\frac{M_{i+1, i+1+k} - M_{i, i+k} }{L}
\,\equiv \,\mu_{1,k}(x_k)
\ee
where the subindex $1$ means that these quantities are related to the first derivative of the functions $\mu_k(x)$.
Taking $k_\textrm{\tiny max} \to \infty$ in (\ref{M-summation k^2 der1 continuum}), we find 
\be
\label{M-summation k^2 der1 continuum infty}
\mathcal{M}_{1,k_\textrm{\tiny max}}^{(2)}(x) \,\longrightarrow\, \mathcal{M}_{1,\infty}^{(2)}(x)\,.
\ee

 \begin{figure}[t!]
\vspace{.2cm}
\hspace{-1.3cm}
% \begin{center}
\includegraphics[width=1.12\textwidth]{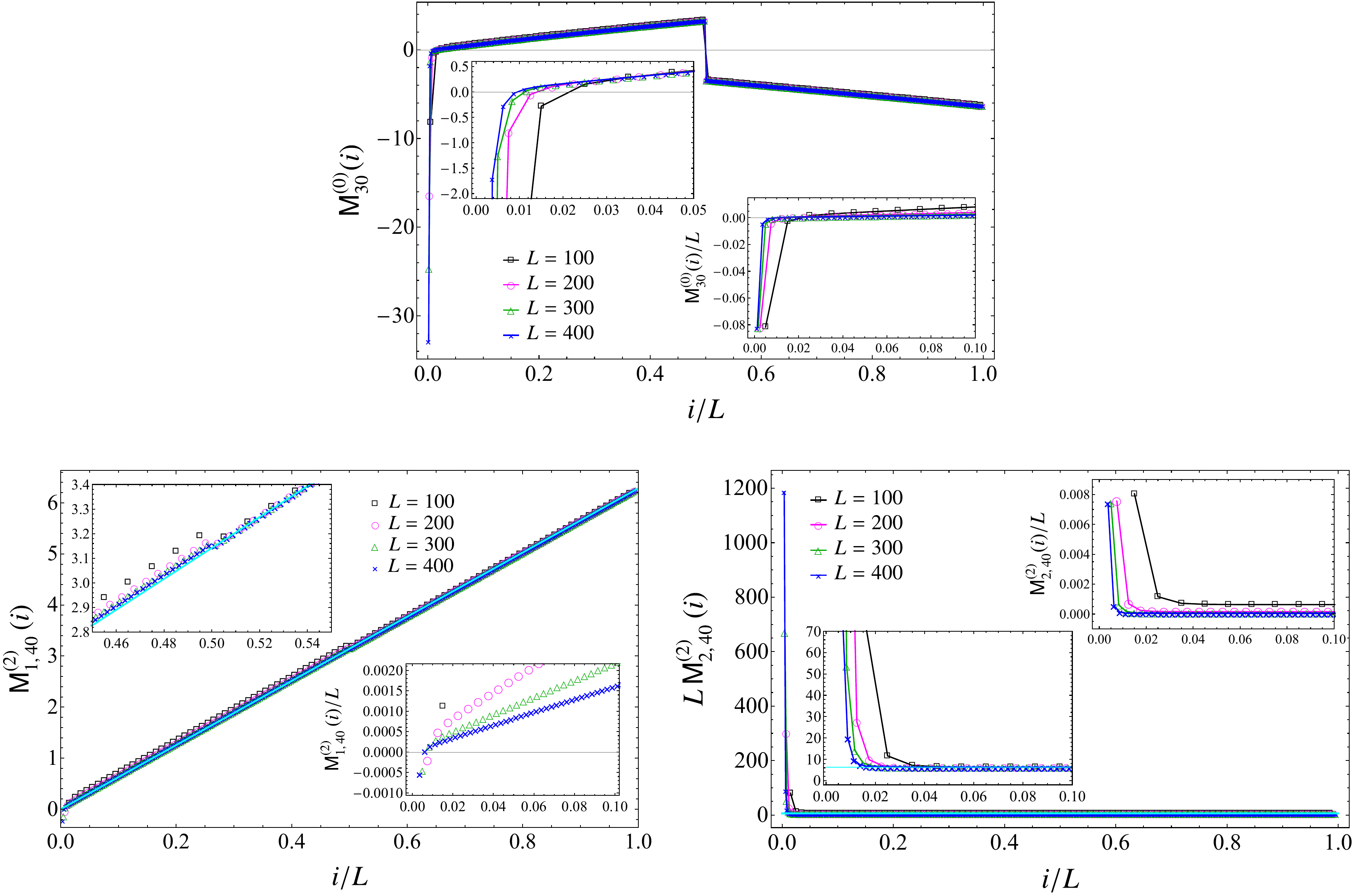}
% \end{center}
\vspace{-.5cm}
\caption{
The combinations (\ref{M-summation k^0 symm}) (top),
(\ref{M-summation k^2 der1 symm}) (bottom left)
and (\ref{M-summation k^2 der2 symm}) (bottom right)
when the subsystem is an interval made by $L$ sites 
at the beginning of the semi-infinite line and $\omega=0$.
The cyan line in the bottom left panel corresponds to $2\pi (i/L)$, while the cyan horizontal line in the bottom right panel corresponds to $2\pi$.
The collapses of the data points for increasing values of $L$ support (\ref{zero-M-functions-bdy}). 
}
\vspace{.1cm}
\label{fig:Zeros-Semi-Infinite}
\end{figure}

Notice that we can also follow the steps performed in \S\ref{sec:infinite-line-massless}
combining the last two terms within the square brackets in (\ref{Moperator expansion1 bdy})
into $ \partial_x [ \mu_k (x) \, \Phi(x)']$ and integrating by parts the corresponding integral,
which provides the boundary terms $[\mu_k(x)  \, \Phi'(x)\, \Phi(x) ] |^{x=\ell}_{x=0}\,$.
These terms do not contribute because $\mu_k(\ell) =0$ at the entangling point
and the Dirichlet boundary condition $\Phi(0) = 0$ holds at the beginning of the semi-infinite line.

Taking the limit $k_{\textrm{\tiny max}} \to \infty$ in (\ref{Moperator expansion1 bdy}) and employing the weight functions introduced in 
(\ref{MN-summation k^0 continuum infty}), (\ref{M-summation k^2 def kmax-infty})
and (\ref{M-summation k^2 der1 continuum infty}),
for the non vanishing contributions to the continuum limit of the entanglement hamiltonian we find
\bea
\label{MplusNoperator expansion bdy}
& & \hspace{-2.5cm}
\frac{H_M + H_N}{2}
=
\frac{\ell}{a^2}
\int_0^\ell 
\frac{1}{2}
\left[ \,\mathcal{M}_{\infty}^{(0)}(x) 
+  \frac{1}{4}\, \mathcal{M}_{2,\infty}^{(2)}(x) \right] 
\Phi(x)^2\, dx\,
\\
\rule{0pt}{.8cm}
& & \hspace{-.1cm}
+\, \ell
\int_0^\ell 
\frac{1}{2}\,
\Big[ \, 
\mathcal{N}_{\infty}^{(0)}(x) \,  \Pi(x)^2 
+ \mathcal{M}_{1,\infty}^{(2)}(x)\, \Phi'(x) \, \Phi(x)
+ \mathcal{M}_{\infty}^{(2)}(x) \, \Phi''(x) \, \Phi(x)
\,\Big] dx\,.
\nonumber
\eea
We remark that, although some formal expressions occur also 
in the case of the interval in the infinite line in \S\ref{sec:infinite-line-massless},
their values depend on the system that we are exploring through 
the correlators (\ref{corr qq dirichlet thermo}) and (\ref{corr pp dirichlet thermo}).

Also in the numerical analysis of this bipartition we have employed all the decompositions 
introduced in \S\ref{sec:EHinHC} and in  Appendix\;\ref{sec_app:details} as starting point.
We find that the most effective approach is based on
(\ref{M-summation k^0 symm}), 
(\ref{N-summation k^0 symm}), (\ref{M-summation k^2 symm})
and (\ref{M-summation k^2 der2 symm}).
For the interval at the beginning of the semi-infinite line, 
we also need the combination of the matrix elements of $M$
for $\mathsf{M}_{1,k_\textrm{\tiny max}}^{(2)}$ and
it is not difficult to find that it reads
\be
\label{M-summation k^2 der1 symm}
\mathsf{M}_{1,k_\textrm{\tiny max}}^{(2)}
=
\begin{cases}
 \sum_{k =1}^{k_{\textrm{\tiny max}}} k^2 
 \big(
 M_{i+1, i+1+k} - M_{i, i+k} \big) & \quad 1\leqslant i\leqslant L/2
 \\
 \rule{0pt}{.7cm}
 \sum_{k =1}^{k_{\textrm{\tiny max}}} k^2 
 \big(
 M_{i-k+1, i+1} - M_{i-k, i}
 \big) & \quad L/2+1\leqslant i\leqslant L\,.
 \end{cases}
\ee
Like for the interval in the infinite line,
the occurrence of two branches in 
(\ref{M-summation k^0 symm}), (\ref{N-summation k^0 symm}), (\ref{M-summation k^2 symm}) 
(\ref{M-summation k^2 der2 symm}) and (\ref{M-summation k^2 der1 symm}) allows to probe the entire interval.
This cannot be done when the decompositions 
(\ref{H_M asymmetric}) and (\ref{H_N asymmetric}) are employed
(see top panels of Fig.\,\ref{fig:sum-methods-M-SemiInfinite} and Fig.\,\ref{fig:sum-methods-N-SemiInfinite}). 
Notice that, in contrast with \S\ref{sec:infinite-line-massless},
in this case the reflection symmetry 
with respect to the center of the interval is not  expected.

\begin{figure}[t!]
\vspace{.2cm}
\hspace{-1.3cm}
% \begin{center}
\includegraphics[width=1.12\textwidth]{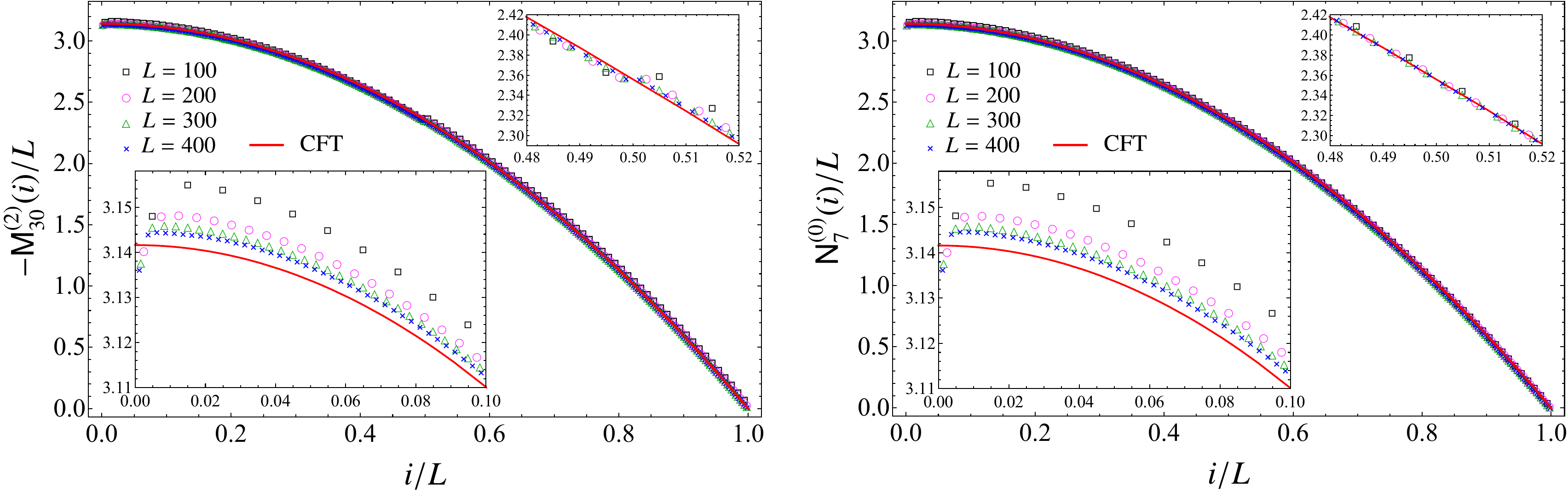}
% \end{center}
\vspace{-.5cm}
\caption{
The combinations (\ref{M-summation k^2 symm}) (left) and (\ref{N-summation k^0 symm}) (right)
when the subsystem is an interval made by $L$ sites at the beginning of the semi-infinite line 
and $\omega=0$.
The collapses of the data points shown support (\ref{MN-beta-functions-bdy}),
with $\beta(x)$ given by the half parabola (\ref{parabola-CFT-semi-infinite})
(red solid curve).
}
\vspace{.0cm}
\label{fig:Parabola-Semi-Infinite}
\end{figure}

In Fig.\,\ref{fig:Zeros-Semi-Infinite} we observe that 
the numerical data for $ \mathsf{M}_{k_\textrm{\tiny max}}^{(0)}(i)$, 
$\mathsf{M}_{1,k_\textrm{\tiny max}}^{(2)}(i)$ and $L \,\mathsf{M}_{2,k_\textrm{\tiny max}}^{(2)}(i)$
with $i \neq 1$ collapse on well defined curves when $L$ increases.
As for some of the weight functions occurring in (\ref{MplusNoperator expansion bdy}), 
these collapses support the following conjecture 
\be
\label{zero-M-functions-bdy}
\mathcal{M}_{\infty}^{(0)}(x) = 0
\;\;\qquad \;\;
\mathcal{M}_{1,\infty}^{(2)}(x) = 0
\;\;\qquad \;\;
\mathcal{M}_{2,\infty}^{(2)}(x) = 0
\ee
for any fixed value of $x \in A$ such that $x \neq 0$.
The insets on the right in all the panels of  Fig.\,\ref{fig:Zeros-Semi-Infinite}
highlight that the values of $\mathsf{M}_{k_\textrm{\tiny max}}^{(0)}(i)/ L$, 
$\mathsf{M}_{1,k_\textrm{\tiny max}}^{(2)}(i) / L$ and
$\mathsf{M}_{2,k_\textrm{\tiny max}}^{(2)}(i) / L$ 
for $i=1$ seem to converge to finite non vanishing constants.
Since $\mathcal{M}_{\infty}^{(0)}(0)$, $\mathcal{M}_{1,\infty}^{(2)}(0)$  and $\mathcal{M}_{2,\infty}^{(2)}(0)$ 
are multiplied by $\Phi(0)$ in (\ref{MplusNoperator expansion bdy}),
the Dirichlet boundary condition $\Phi(0) = 0$
implies that this feature does not provide a non vanishing term in the continuum limit of the entanglement hamiltonian.
%%%
In the top panel of Fig.\,\ref{fig:Zeros-Semi-Infinite},
the discontinuity in the center of the interval 
is due to the fact that $\mathsf{M}_{k_\textrm{\tiny max}}^{(0)}$  
in (\ref{M-summation k^0 symm}) is defined through two branches.
This discontinuity is not observed if different decompositions 
for the operators $\widehat{H}_M$ and $\widehat{H}_N$ in (\ref{H_M and H_N operators})
are adopted.

In Fig.\,\ref{fig:Parabola-Semi-Infinite} we show 
$-\mathsf{M}_{k_\textrm{\tiny max}}^{(2)}/L$ (left panel) and $\mathsf{N}_{k_\textrm{\tiny max}}^{(0)}/L$ (right panel)
for increasing values of $L$ and an optimal value of $k_{\textrm{\tiny max}}$ which guarantee certain stability of the numerical results. 
The collapses of the data points naturally lead to conjecture that
\be
\label{MN-beta-functions-bdy}
 \mathcal{M}_{\infty}^{(2)}(x) 
 = -\, \beta(x)
%= -\, 2\pi \, \frac{x}{\ell} \left( 1-\frac{x}{\ell} \right)
\hspace{.6cm} \qquad \hspace{.6cm}
 \mathcal{N}_{\infty}^{(0)}(x) 
 =  \beta(x)
%= 2\pi \, \frac{x}{\ell} \left( 1-\frac{x}{\ell} \right)
\ee
where $\beta(x)$ is the half parabola (\ref{parabola-CFT-semi-infinite}) predicted by the CFT.
Comparing these results with the corresponding ones for the interval in the infinite line (see Fig.\,\ref{fig:Parabola-Infinite}), 
we observe that larger values of $L$ are needed in this case 
to reach the CFT curve in the neighbourhood of the beginning of the semi-infinite line,
which is sensible to the boundary conditions. 

 \begin{figure}[t!]
\vspace{.2cm}
\hspace{-1.3cm}
% \begin{center}
\includegraphics[width=1.12\textwidth]{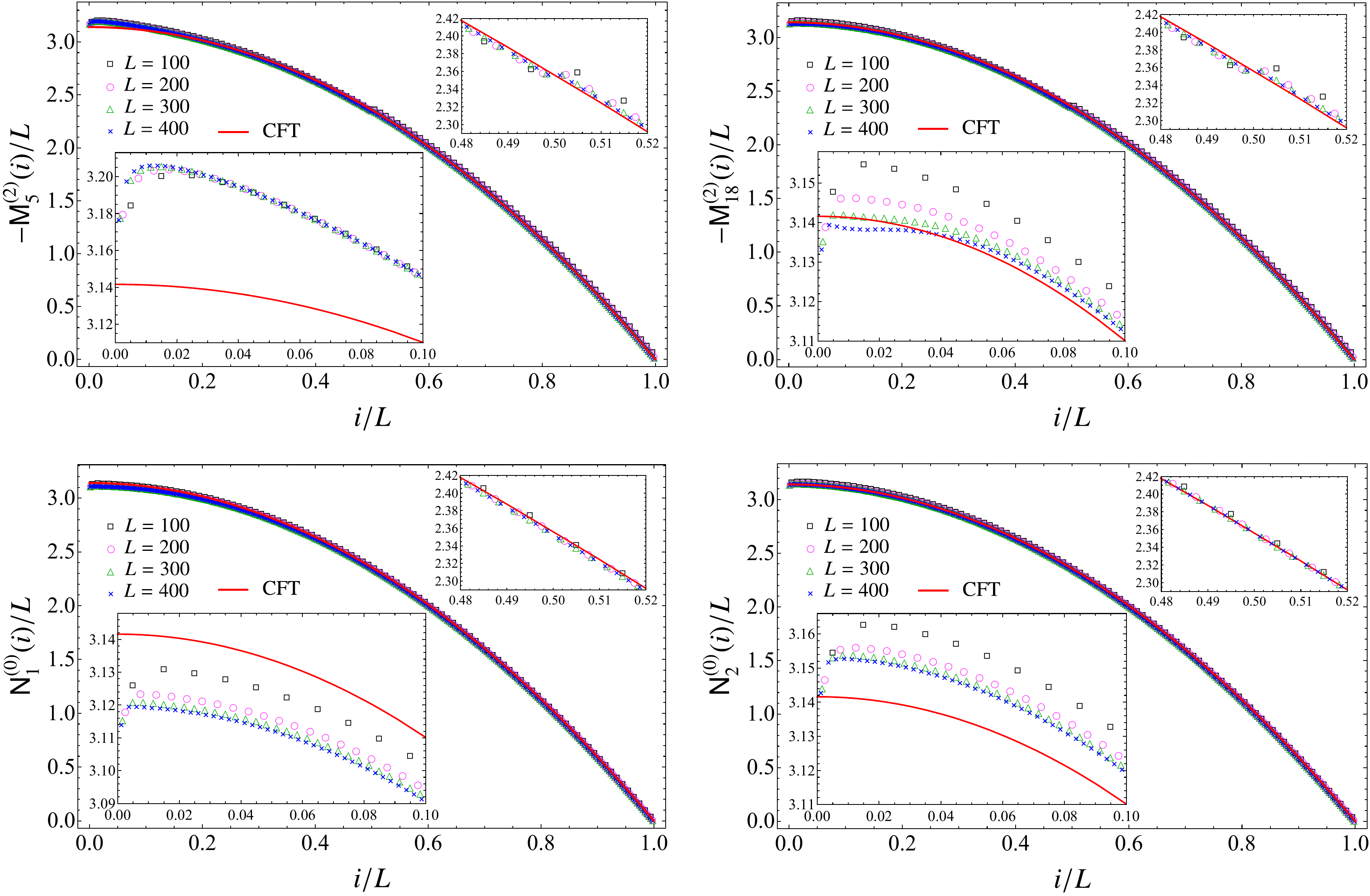}
% \end{center}
\vspace{-.5cm}
\caption{
Role of the parameter $k_{\textrm{\tiny max}}$
in the combinations (\ref{M-summation k^2 symm}) (top panels)
and (\ref{N-summation k^0 symm}) (bottom panels)
when the subsystem is an interval made by $L$ sites 
at the beginning of the semi-infinite line 
and $\omega=0$.
The insets, which zoom in on the left endpoint and on the central part of the interval,
show that the agreement with the CFT prediction
given by the half parabola (\ref{parabola-CFT-semi-infinite}) (red solid curve)
improves as $k_{\textrm{\tiny max}}$ increases.
}
\vspace{.0cm}
\label{fig:kmax-Symm-SemiInfinite}
\end{figure}

In the bottom panels of Fig.\,\ref{fig:Zeros-Semi-Infinite},
the data points for $\mathsf{M}_{1,k_\textrm{\tiny max}}^{(2)}$ 
and $L \,\mathsf{M}_{2,k_\textrm{\tiny max}}^{(2)}(i)$ 
with $i \neq 1$ collapse on the cyan straight lines,
which correspond respectively to $2 \pi \, x/\ell = -\ell \beta'(x)$
and to $2\pi= -\ell^2 \beta''(x)$ when $L$ is large enough. 
Similarly to the case of the interval in the infinite line 
(see the final remarks of \S\ref{sec:infinite-line-massless}),
we can roughly justify this behaviour by noticing that
$\mathsf{M}_{1,k_\textrm{\tiny max}}^{(2)}$ and $\mathsf{M}_{2,k_\textrm{\tiny max}}^{(2)}$
are obtained through finite differences approximations of 
$\mu_k'(x)$ and $\mu_k''(x)$ respectively;
hence, from (\ref{MN-beta-functions-bdy}), 
one expects to find respectively $-\beta(x)'$ and $-\beta(x)''$.
Also in this case exchanging
the derivatives with respect to $x$ with the discrete sums over $k$ leads to wrong results,
as already discussed in the final part of \S\ref{sec:infinite-line-massless} for the interval in the infinite line.

In Fig.\,\ref{fig:kmax-Symm-SemiInfinite} we show again 
$-\mathsf{M}_{k_\textrm{\tiny max}}^{(2)}/L$ and $\mathsf{N}_{k_\textrm{\tiny max}}^{(0)}/L$,
but for lower values of $k_\textrm{\tiny max}$ in order to highlight 
the fact that the collapse of the numerical data onto the CFT curve 
improves as $k_\textrm{\tiny max}$ increases.
This behaviour is stabilised around optimal values for  $k_\textrm{\tiny max}$
that correspond to the data reported in Fig.\,\ref{fig:Parabola-Semi-Infinite}.
Furthermore, also in this case we encounter the same parity effect observed in 
Fig.\,\ref{fig:kmax-Symm-Infinite} and mentioned in \S\ref{sec:infinite-line-massless}.

%%%%%%%%%%%%%%%%%%%%%%%%%%%%%%%%%%%%%%%%%%%%%%%%%%%%%%%%
%%%%%%%%%%%%%%%%%%%%%%%%%%%%%%%%%%%%%%%%%%%%%%%%%%%%%%%%
\subsection{Entanglement spectrum}
\label{sec:semi-infinite-line-gaps}

The BCFT analysis of the entanglement spectrum presented in \cite{ct-16},
where (\ref{ratios_bcft_intro}) has been derived, 
includes the case that we are considering, 
given by the entire system in its ground state and the interval $A$ at the beginning of the semi-infinite line. 
In this bipartition only one entangling point occurs; hence the UV cutoff $\epsilon$ is
introduced by removing only a disk of radius $\epsilon$ around the entangling point.
The resulting euclidean spacetime has the topology of the annulus and in this case
different conformal boundary conditions are allowed at the two boundaries. 
In our analysis we impose Dirichlet boundary conditions along the boundary corresponding
to the beginning of the semi-infinite line.

\begin{figure}[t!]
\vspace{.2cm}
\hspace{-1.3cm}
%\begin{center}
\includegraphics[width=1.12\textwidth]{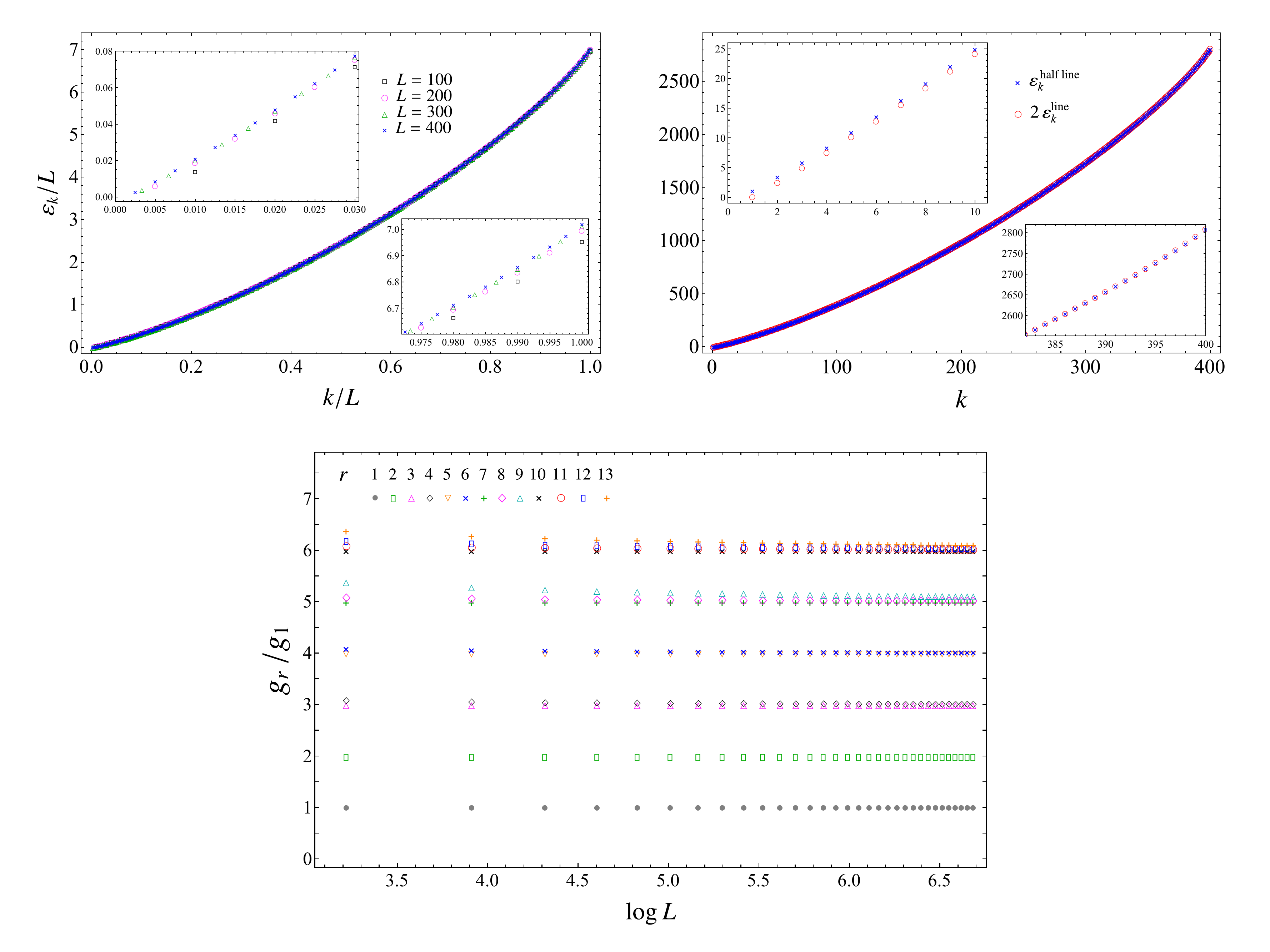}
%\end{center}
\vspace{-.5cm}
\caption{
Entanglement spectrum for an interval at the beginning of the semi-infinite line with Dirichlet boundary conditions.
Top left: The single particle entanglement energies $\varepsilon_k$ for different values of $L$
(the insets zoom in on the lowest and on the highest values of $k$).
Top right: Comparison between the single particle entanglement energies $\varepsilon_k$ 
for an interval with $L=400$ sites at the beginning of the semi-infinite line with Dirichlet boundary conditions
(see the top left panel)
and the ones for an interval with $L=400$ sites  in the infinite line
(see the top right panel of Fig.\,\ref{fig:gaps-Infinite}).
Bottom:
The ratios of the gaps in the entanglement spectrum as functions of $\log L$ when $\omega =0$.
}
\vspace{.0cm}
\label{fig:gaps-Semi-Infinite}
\end{figure}

The numerical analysis of the entanglement spectrum is performed like in  \S\ref{sec:infinite-line-gaps}
and the crucial difference with respect to the interval in the infinite line is that 
the massless regime given by $\omega = 0$ is well defined.
In this regime we observe that the lowest single particle entanglement energy $\varepsilon_1$ is non vanishing,
in contrast with the case of the interval in the infinite line.

In the top left panel of Fig.\,\ref{fig:gaps-Semi-Infinite} we show the numerical results for the single particle entanglement energies 
$\varepsilon_k/L$ in terms of $k/L$ corresponding to some values of $L$, finding that they nicely collapse on a well defined curve
when $L$ is large enough, like in the case of the interval in the infinite line (see the top right panel of Fig.\,\ref{fig:gaps-Infinite}). 
The curves obtained for these two spatial bipartitions are compared in the top right panel of Fig.\,\ref{fig:gaps-Semi-Infinite},
finding that they basically overlap, once the curve for the interval in the infinite line is multiplied by a factor of $2$
(the insets highlight that this agreement is very good in the highest part of the spectrum and gets worse in the lowest part of the spectrum).

In the bottom panel of Fig.\,\ref{fig:gaps-Semi-Infinite} we show the ratios $g_r/g_1$ 
between the generic gap $g_r$ and the smallest gap $g_1$
in the entanglement spectrum as functions of $\log L$, for $1\leqslant r \leqslant 13$.
These ratios take all the integer values between $1$ and $6$ included
(we checked that $g_r/g_1 > 6.5$ for $r>13$ for the largest value of $L$ at our disposal). 
This feature originates from the fact that in the low-lying part of the single particle 
entanglement spectrum the eigenvalues are equally separated by an integer multiple of $\varepsilon_1$
(see also \S\ref{sec:infinite-line-gaps}).
Comparing Fig.\,\ref{fig:gaps-Semi-Infinite} and Fig.\,\ref{fig:gaps-Infinite},
it is straightforward to notice that  $g_r/g_1$ take all the integer values 
for both the bipartitions, but the corresponding degeneracies are very different in the two cases. 
In particular, in Fig.\,\ref{fig:gaps-Semi-Infinite}
the degeneracy of the $n$-th level is given by 
the number of partitions of $n$ that do not contain repeated positive integers.

The degeneracy observed in  Fig.\,\ref{fig:gaps-Semi-Infinite}
is compatible with the conformal spectrum of the free massless scalar 
on a segment with mixed boundary conditions, namely
with Dirichlet boundary conditions imposed at one endpoint 
and Neumann boundary conditions at the other endpoint \cite{blumenhagen}.
Since in our analysis Dirichlet boundary conditions are imposed at
the beginning of the semi-infinite line, we can conclude that
Neumann boundary conditions must be imposed at the boundary introduced
by the regularisation procedure around the entangling point. 
This allows to fix the ambiguity found in \S\ref{sec:infinite-line-gaps},
concluding that the numerical results for the entanglement spectrum
of the interval in the infinite line in the continuum limit agree with 
the conformal spectrum of the BCFT given by the free massless scalar 
on a segment with Neumann boundary conditions imposed 
on both the boundaries encircling the endpoints of the interval in the euclidean spacetime. 
Also for this bipartition we expect that the agreement with the conformal spectrum of the BCFT
holds only for the low-lying part of the entanglement spectrum.

%\newpage
%%%%%%%%%%%%%%%%%%%%%%%%%%%%%%%%%%%%%%%%%%%%%%%%%
%%%%%%%%%%%%%%%%%%%%%%%%%%%%%%%%%%%%%%%%%%%%%%%%%
\section{Conclusions}
\label{sec:conclusions}

In this manuscript we have performed a numerical analysis of the
continuum limit of the entanglement hamiltonians of a block made by $L$ 
consecutive sites in massless harmonic chains,
in the two cases where the subsystem is an interval in the infinite line 
or an interval at the beginning of the semi-infinite line 
with Dirichlet boundary conditions imposed at its endpoint.
The procedure is based on the method introduced in \cite{ep-17, etp-19} for 
chains of free fermions, which has been adapted here to harmonic chains. 
%%%

We have obtained the analytic expression (\ref{EH-intro}) predicted by CFT,
with the weight function $\beta(x)$ and the energy density $T_{00}(x)$
respectively given by (\ref{parabola-CFT-infinite}) and (\ref{T00 massless infinite}) for the interval in the infinite line
and by (\ref{parabola-CFT-semi-infinite})  and (\ref{T00 massless semi-infinite}) 
for the interval at the beginning of a semi-infinite line.
A remarkable agreement between the data points and the weight functions $\beta(x)$ predicted by CFT
is observed (see Fig.\,\ref{fig:Parabola-Infinite} and Fig.\,\ref{fig:Parabola-Semi-Infinite}).
%%%
It would be instructive to support our numerical results with analytic computations,
by first finding analytic expressions for the functions $\mu_k(x)$ and $\nu_k(x)$
(see Fig.\,\ref{fig:Muk-Infinite}, Fig.\,\ref{fig:Nuk-Infinite} for the interval in the infinite line and 
Fig.\,\ref{fig:Muk-Semi-Infinite} and Fig.\,\ref{fig:Nuk-Semi-Infinite} 
for the interval at the beginning of the semi-infinite line) 
and then by performing analytically the sums involving these functions and their derivatives
which provide continuum limit of the entanglement hamiltonians, 
as done in \cite{ep-17, etp-19}  for the interval in the infinite chain of free fermions.

We have also explored the continuum limit of the entanglement spectra of these entanglement hamiltonians,
finding that the ratios of the low-lying gaps  provide the ratios of the conformal dimensions of 
the BCFT given by the massless scalar on the annulus with 
the proper conformal boundary conditions, as predicted in \cite{ct-16}
 (see Fig.\,\ref{fig:gaps-Infinite} and Fig.\,\ref{fig:gaps-Semi-Infinite}).
The numerical results indicate that Neumann boundary conditions 
must be imposed along the boundaries introduced by the regularisation procedure.
This is in agreement with a similar numerical analysis performed in lattice spin models
\cite{lauchli-spectrum}, where it has been found that the numerical results for the entanglement spectra
are compatible with the conformal spectra of BCFT with
free boundary conditions imposed along the boundaries around the entangling points.
This has been confirmed also by numerical studies out of equilibrium \cite{stt-19}.

The results reported in this manuscript can be extended in various directions.
In massless harmonic chains, the entanglement hamiltonians of an interval 
in a circle when the system is in its ground state
or in the infinite line when the system is at finite temperature should be studied
because in these cases (\ref{EH-intro}) still holds and the weight functions $\beta(x)$ are known from CFT 
\cite{klich-13, ct-16, ep-18, etp-19}.
It is also natural to explore the entanglement hamiltonians for bipartitions involving 
disjoint intervals \cite{disjoint intervals cft, ch-09-eh-2int, Arias-18},
spatially inhomogeneous chains \cite{dubail-curved, trs-18-rainbow} 
and higher dimensional quantum systems \cite{chm}. 
It is important to find explicit expressions for the entanglement hamiltonians
in interacting lattice models, both through analytic 
and numerical methods \cite{peschel-truong, parisentoldin, calabrese-campostrini, ludwig-ryu-ESgapped-17, klich-wong}.
%%%
Also the analysis of the entanglement spectra
\cite{lauchli-spectrum, calabrese-lefevre, alba-lauchli-ESgapped-12, 
alba-calabrese-tonni, assaad-ent-spectrum, sierra-cirac-18} and of the 
contour for the entanglement entropies \cite{br-04, chen-vidal, cdt-17-contour},
which is a quantity strictly related to the entanglement hamiltonian,
deserve further analysis. 
It is useful also to study operators on the lattice that provides efficient approximations
of the entanglement hamiltonians  \cite{trs-18-rainbow, dalmonte}.
%%%
We also mention that,
in order to understand the unitary time evolution of a 
system after a quantum quench, 
i.e. after a sudden change that drives the system out of equilibrium \cite{cc-quench},
relevant insights could come from the analysis of the time evolutions of 
the entanglement hamiltonians
and of their entanglement spectra 
\cite{ct-16, tagliacozzo-torlai, dat-19, stt-19, Wen- eh quench 19, ryu-ludwig-18-cft}.

%%%%%%%%%%%%%%%%%%%%%%%%%%%%%%%%%%%%%%%%%%%%%%%%
%\newpage

\section*{\small Acknowledgements}

It is our pleasure to thank Ra\'ul Arias, Viktor Eisler, Mihail Mintchev and Ingo Peschel 
for important comments. 
We are grateful also to Vincenzo Alba, John Cardy, Paul Fendley,  Andreas Ludwig, 
Giuseppe Mussardo, German Sierra, Jacopo Surace and Luca Tagliacozzo for useful discussions.
ET acknowledges the Yukawa Institute for Theoretical Physics at Kyoto University
(workshop YITP-T-19-03 {\it Quantum Information and String Theory 2019})
and the Instituto de F\'isica Te\'orica (Madrid)
for financial support and warm hospitality during part of this work.

\newpage
%%%%%%%%%%%%%%%%%%%%%%%%%%%%%%%%%%%%%%%%%%%%%%%%%%%%%%%

\begin{appendices}

\section*{Appendices}

%%%%%%%%%%%%%%%%%%%%%%%%%%%%%%%%%%%%%%%%%%%%%%%%%%%%%%%
%%%%%%%%%%%%%%%%%%%%%%%%%%%%%%%%%%%%%%%%%%%%%%%%%%%%%%%
\section{Alternative summations and role of $k_{\textrm{\tiny max}}$}
\label{sec_app:details}

In this Appendix we report further results supporting the numerical analysis 
discussed in the main text. 
First we briefly discuss the choice of the numerical value of $\omega L$ 
adopted to study the entanglement hamiltonian of the interval in the infinite line. 
In the remaining part of this Appendix, 
we discuss some numerical results obtained through 
decompositions of the operators 
$\widehat{H}_M$ and $\widehat{H}_N$ in (\ref{H_M and H_N operators}) 
that are different from (\ref{H_M symm}) and (\ref{H_N symm}).

\begin{figure}[t!]
\vspace{.2cm}
\hspace{-1.3cm}
% \begin{center}
\includegraphics[width=1.12\textwidth]{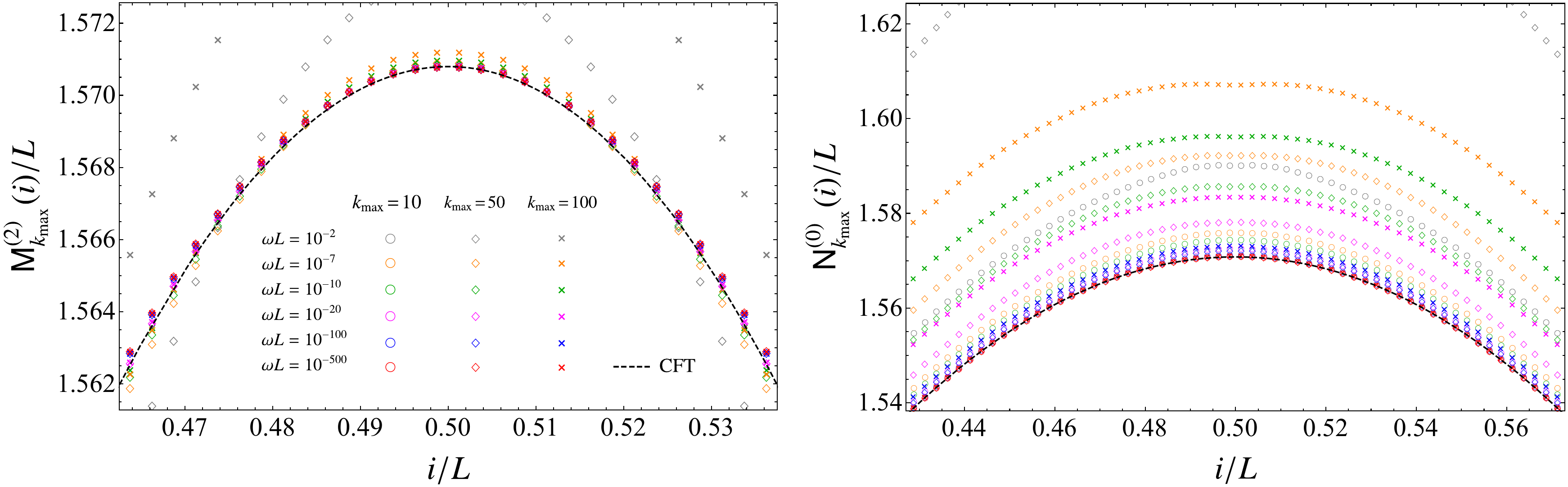}
% \end{center}
\vspace{-.5cm}
\caption{
The combination (\ref{M-summation k^2 symm}) (left) and  the combination (\ref{N-summation k^0 symm}) (right) 
for decreasing values of $\omega L\ll 1$ and increasing values of $k_{\textrm{\tiny max}}$
when the subsystem is an interval made by $L=400$ sites in the infinite chain. 
The dashed black line corresponds to the parabola (\ref{parabola-CFT-infinite}) predicted by CFT.
}
\vspace{.0cm}
\label{fig:zero-mass}
\end{figure}

The harmonic chain on the infinite line displays translation invariance
and this symmetry leads to the occurrence of the zero mode, 
that prevents us to set $\omega = 0$ in our numerical analysis,
as already remarked in \S\ref{sec:infinite-line-corr}.
In order to study the entanglement hamiltonians in the massless regime of this harmonic chain,
we have to choose very small but non vanishing values of $\omega >0$.
The high numerical precision required for our numerical analysis 
allows to take $\omega >0$ significantly close to zero.
The numerical data reported in the figures discussed in 
\S\ref{sec:infinite-line-massless} and \S\ref{sec:infinite-line-gaps}
correspond to $\omega L = 10^{-500}$
and in Fig.\,\ref{fig:zero-mass} we justify this choice by showing 
$\mathsf{M}_{k_\textrm{\tiny max}}^{(2)}/L$ and $\mathsf{N}_{k_\textrm{\tiny max}}^{(0)}/L$ 
for decreasing values of $\omega L$ and for three increasing values of $k_{\textrm{\tiny max}}$ at fixed $L=400$.
In order to find a small value for $\omega L >0$ that properly captures the features of the massless regime,
we require that the numerical data are stabilised on the CFT prediction (\ref{parabola-CFT-infinite})
for different values of $k_{\textrm{\tiny max}}$.
While this condition is fulfilled already for $\omega L = 10^{-20}$ 
in the left panel of Fig.\,\ref{fig:zero-mass}, it is not satisfied in the right panel. 
Instead, for $\omega L = 10^{-500}$ very good collapses on the CFT curve are observed 
for the data corresponding to different $k_{\textrm{\tiny max}}$.

In the main text we have discussed the results for the continuum limit reported
in Fig.\,\ref{fig:Zeros-Infinite} and Fig.\,\ref{fig:Parabola-Infinite} for the interval in the infinite line
(see \S\ref{sec:infinite-line-massless})
and in Fig.\,\ref{fig:Zeros-Semi-Infinite} and Fig.\,\ref{fig:Parabola-Semi-Infinite} 
for the interval at the beginning of the semi-infinite line
(see \S\ref{sec:semi-infinite-line-massless}).
They have been obtained 
by starting from the decompositions (\ref{H_M symm}) and (\ref{H_N symm}) 
for $\widehat{H}_M$ and $\widehat{H}_N$,
which lead to the combinations
(\ref{M-summation k^0 symm}), (\ref{N-summation k^0 symm}), 
(\ref{M-summation k^2 symm}) and (\ref{M-summation k^2 der2 symm})
for the interval in the infinite line,
and to the same combinations together with (\ref{M-summation k^2 der1 symm})
for the interval at the beginning of the semi-infinite line.

\begin{figure}[t!]
\vspace{.5cm}
\hspace{-1.3cm}
% \begin{center}
\includegraphics[width=1.12\textwidth]{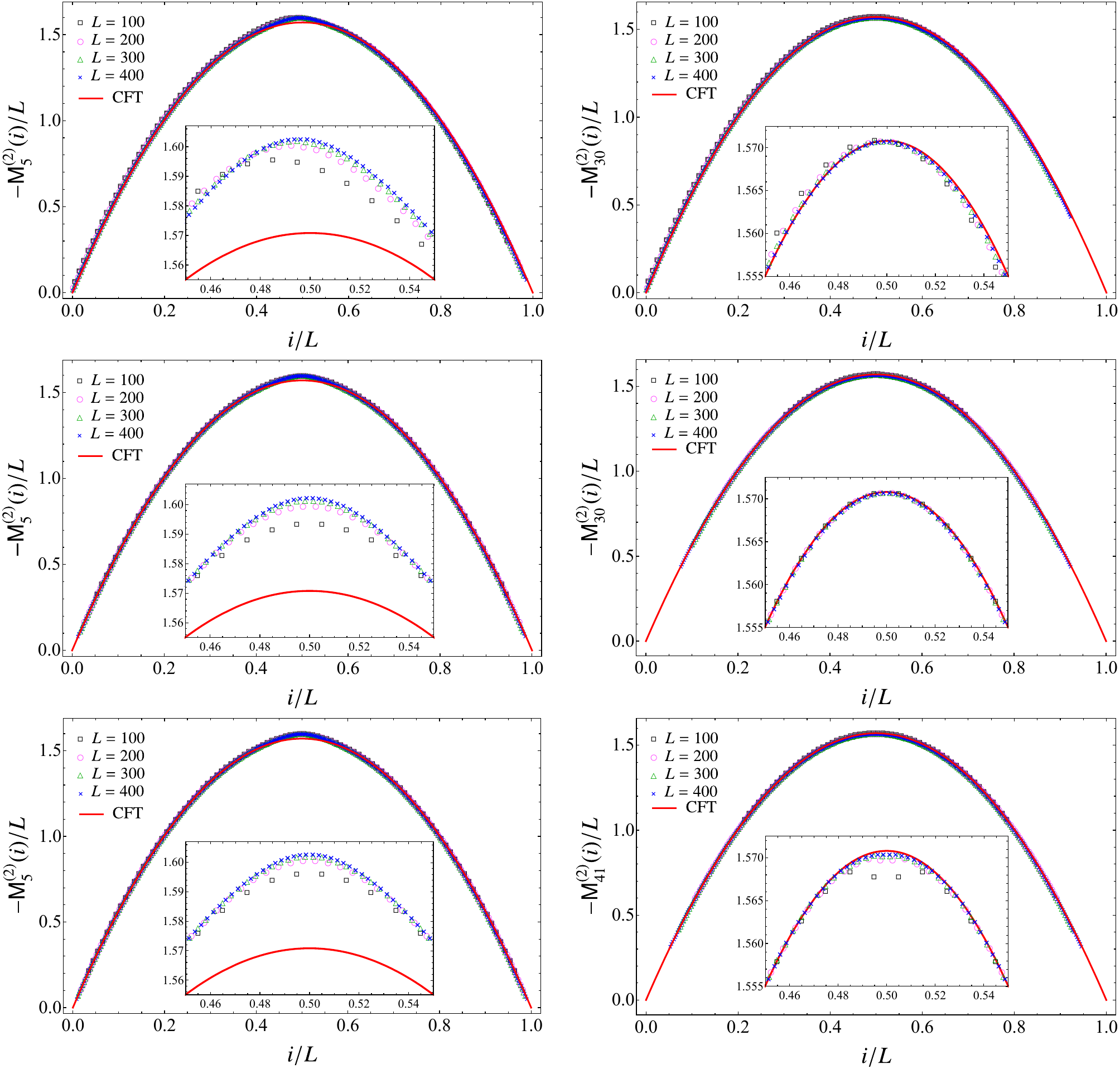}
% \end{center}
\vspace{-.5cm}
\caption{
The combinations given by (\ref{M-summation k^2 def}) (top panels), 
by the first expression in (\ref{M-N-summation Casini}) (middle panels) 
and by (\ref{M-summation k^2 Eisler}) (bottom panels) 
for different values of $L$ and 
two values of $k_{\textrm{\tiny max}}$ for each combination (left and right panels),
when the subsystem is an interval in the infinite line.
The red solid curve is the parabola (\ref{parabola-CFT-infinite}) predicted by CFT.
}
\vspace{.0cm}
\label{fig:sum-methods-M-Infinite}
\end{figure}

\begin{figure}[t!]
\vspace{.5cm}
\hspace{-1.3cm}
% \begin{center}
\includegraphics[width=1.12\textwidth]{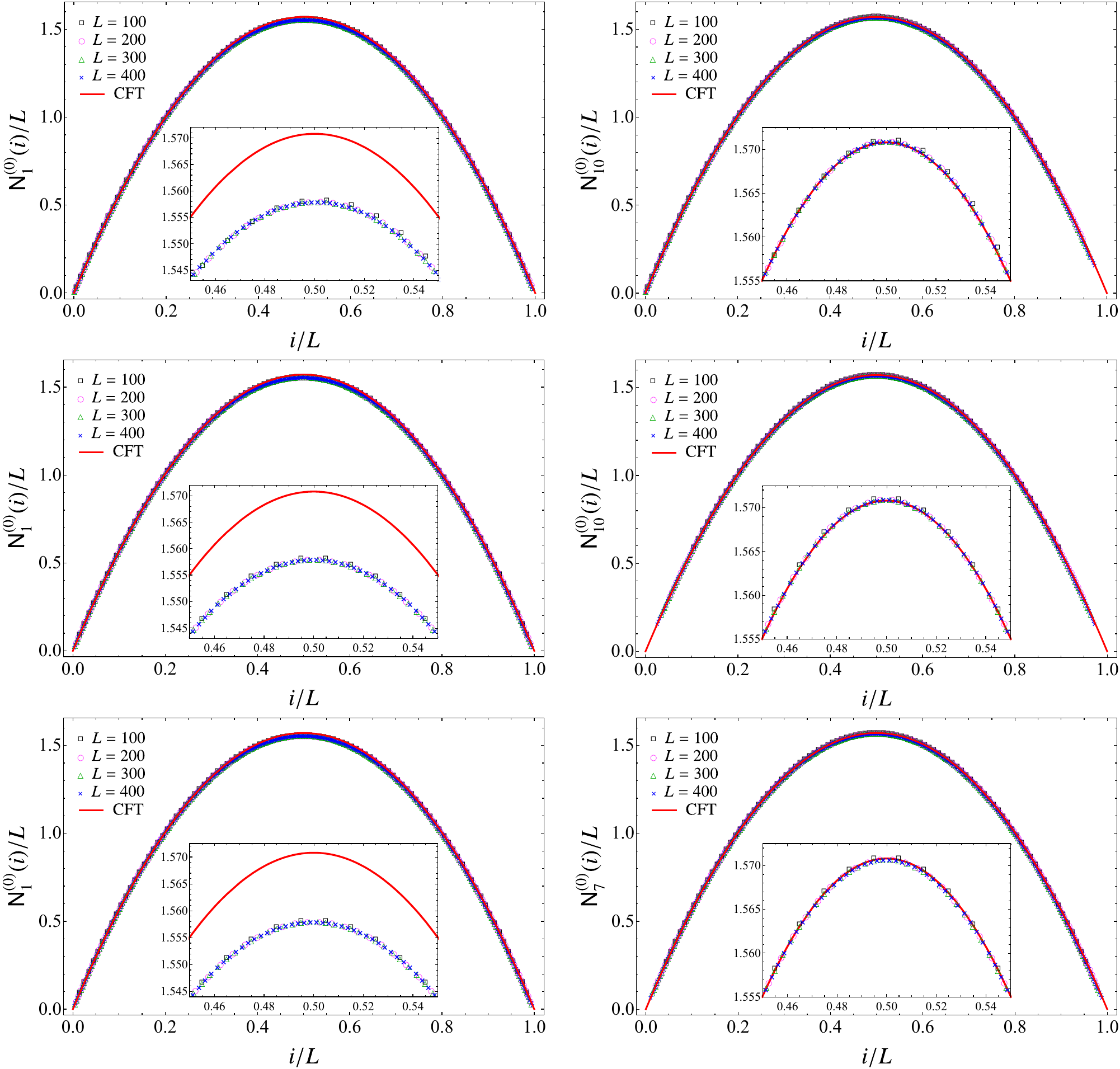}
% \end{center}
\vspace{-.5cm}
\caption{
The combinations given by (\ref{N-summation k^0 def}) (top panels), 
by the second expression in (\ref{M-N-summation Casini}) (middle panels) 
and by (\ref{N-summation k^0 Eisler}) (bottom panels) 
for different values of $L$ and 
two values of $k_{\textrm{\tiny max}}$ for each combination (left and right panels),
when the subsystem is an interval in the infinite line.
The red solid curve is the parabola (\ref{parabola-CFT-infinite}) predicted by CFT.
}
\vspace{.0cm}
\label{fig:sum-methods-N-Infinite}
\end{figure}

\begin{figure}[t!]
\vspace{.5cm}
\hspace{-1.3cm}
% \begin{center}
\includegraphics[width=1.12\textwidth]{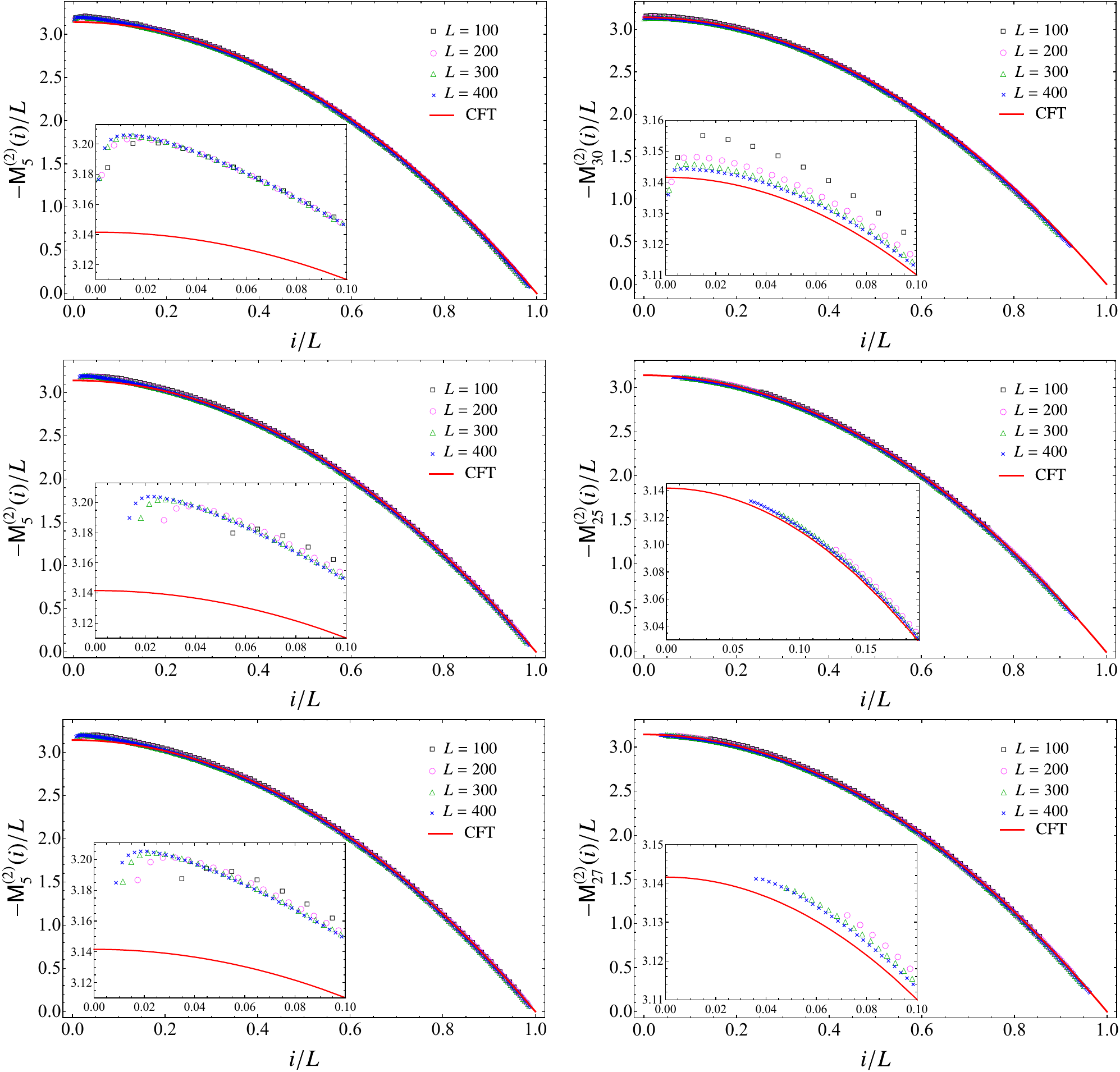}
% \end{center}
\vspace{-.5cm}
\caption{
The combinations given by (\ref{M-summation k^2 def}) (top panels), 
by the first expression in (\ref{M-N-summation Casini}) (middle panels) 
and by (\ref{M-summation k^2 Eisler}) (bottom panels) 
for different values of $L$ and 
two values of $k_{\textrm{\tiny max}}$ for each combination (left and right panels),
when the subsystem is an interval at the beginning of  the semi-infinite line.
The red solid curve is the half parabola (\ref{parabola-CFT-semi-infinite}) predicted by CFT.
}
\vspace{.0cm}
\label{fig:sum-methods-M-SemiInfinite}
\end{figure}

\begin{figure}[t!]
\vspace{.5cm}
\hspace{-1.3cm}
% \begin{center}
\includegraphics[width=1.12\textwidth]{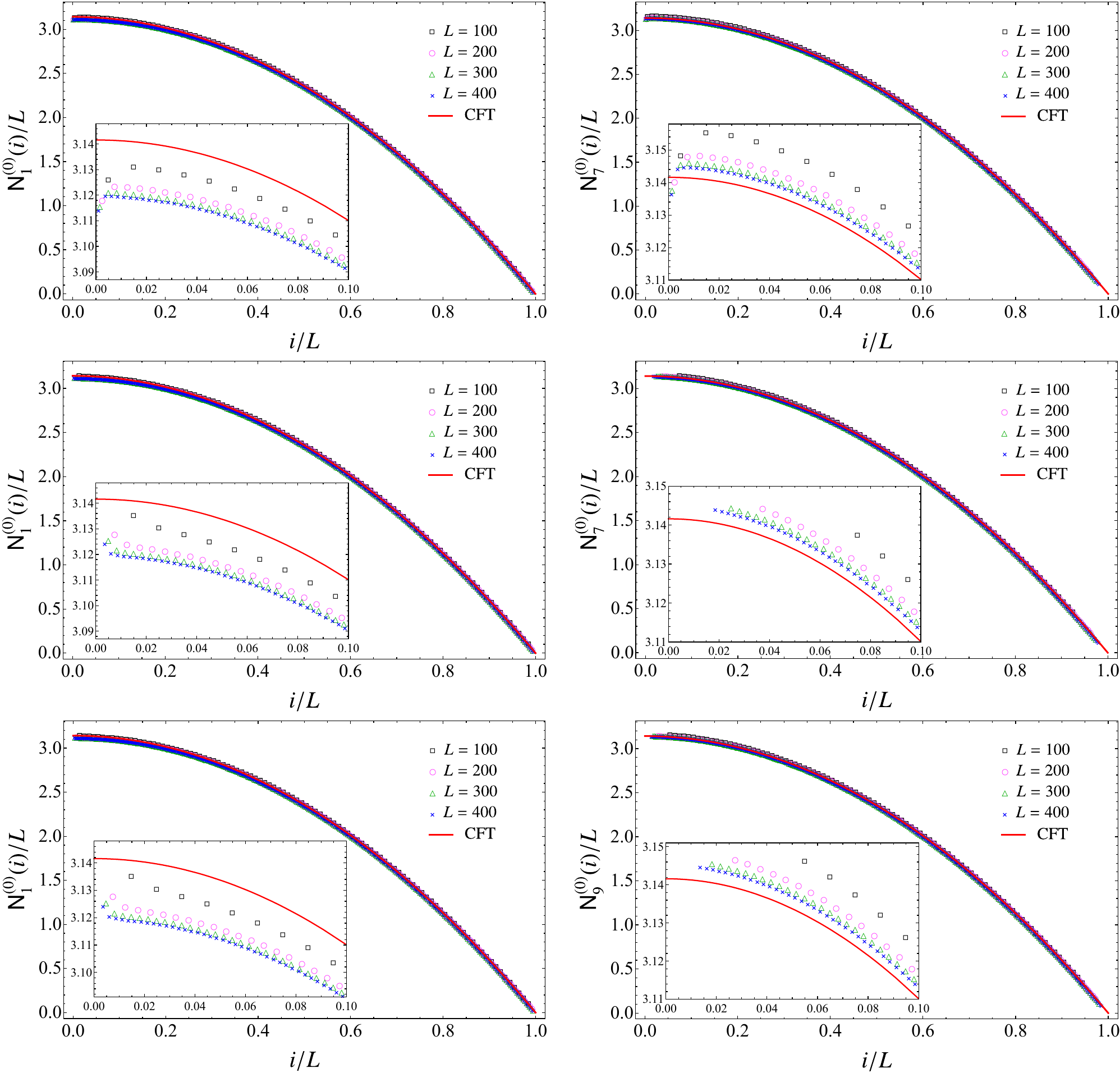}
% \end{center}
\vspace{-.5cm}
\caption{
The combinations given by (\ref{N-summation k^0 def}) (top panels), 
by the second expression in (\ref{M-N-summation Casini}) (middle panels) 
and by (\ref{N-summation k^0 Eisler}) (bottom panels) 
for different values of $L$ and 
two values of $k_{\textrm{\tiny max}}$ for each combination (left and right panels),
when the subsystem is an interval at the beginning of  the semi-infinite line.
The red solid curve is the half parabola (\ref{parabola-CFT-semi-infinite}) predicted by CFT.
}
\vspace{.0cm}
\label{fig:sum-methods-N-SemiInfinite}
\end{figure}

The procedure to study the continuum limit starting from the decompositions
(\ref{H_M asymmetric}) and (\ref{H_N asymmetric}) for $\widehat{H}_M$ and $\widehat{H}_N$,
which provides 
the combinations (\ref{M-summation k^0 def}), (\ref{N-summation k^0 def}),
(\ref{M-summation k^2 def}) and (\ref{M-summation k^2 der2 def})
for the interval in the infinite line (see \S\ref{sec:infinite-line-massless})
and the same combinations together with (\ref{M-summation k^2 der1 def})
for the interval at the beginning of the semi-infinite line
(see \S\ref{sec:semi-infinite-line-massless}),
has been explained in the main text. 
%%%%%
The numerical results of the combinations (\ref{M-summation k^2 def}) and (\ref{N-summation k^0 def})
for various sizes $L$ of the intervals and for two values of $k_\textrm{\tiny max}$ are shown
in the top panels of Fig.\,\ref{fig:sum-methods-M-Infinite} and Fig.\,\ref{fig:sum-methods-N-Infinite}
for the interval in the infinite line
and of Fig.\,\ref{fig:sum-methods-M-SemiInfinite} and Fig.\,\ref{fig:sum-methods-N-SemiInfinite}
for the interval at the beginning of the semi-infinite line.
As $k_\textrm{\tiny max}$ increases, 
the agreement between the data and the CFT predictions 
given by (\ref{parabola-CFT-infinite}) and (\ref{parabola-CFT-semi-infinite}) improves.
The numerical data stabilise around a value that has been adopted in the right panels. 
%%%%%
As for the combinations (\ref{M-summation k^0 def}) and (\ref{M-summation k^2 der2 def}),
we find that they lead to the function that vanishes identically in the interval,
as expected from CFT.
The range of the index $i$ in these combinations do not allow to capture the curve predicted by CFT 
on the entire interval 
and this fact motivated us to employ the decompositions (\ref{H_M symm}) and (\ref{H_N symm})
instead of (\ref{H_M asymmetric}) and (\ref{H_N asymmetric}).

In the remaining part of this Appendix, we discuss 
the continuum limit of the entanglement hamiltonians
based on two other decompositions for 
$\widehat{H}_M$ and $\widehat{H}_N$ in (\ref{H_M and H_N operators}).

Considering the decompositions (\ref{H_M Casini}) and (\ref{H_N Casini}), 
by adapting the procedure described in \S\ref{sec:infinite-line-massless},
we find the combinations given by
\be
\label{M-N-summation Casini}
\mathsf{M}_{k_\textrm{\tiny max}}^{(2)}(i)
\,= \!\!\!
\sum_{k=-k_\textrm{\tiny max}}^{k_\textrm{\tiny max}}\!\! \frac{k^2}{2} 
\,M_{i,i+k}
\;\;\qquad\;\;
\mathsf{N}_{k_\textrm{\tiny max}}^{(0)}(i)
\, =\!\!\!
\sum_{k=-k_\textrm{\tiny max}}^{k_\textrm{\tiny max}} \!\! \!\! N_{i,i+k}
\ee
and
\be
\label{M-summation k^0 and k^2 der Casini}
\mathsf{M}_{k_\textrm{\tiny max}}^{(0)}(i)
\, =\!\!\!
\sum_{k=-k_\textrm{\tiny max}}^{k_\textrm{\tiny max}}  
\!\!\!\! M_{i,i+k}
\;\qquad\;
\mathsf{M}_{2,\,k_\textrm{\tiny max}}^{(2)}(i)
\, =\!\!\!
\sum_{k=-k_\textrm{\tiny max}}^{k_\textrm{\tiny max}}
\!\! \frac{k^2}{2}
\Big( 
M_{i+1, i+1+k} - 2M_{i, i+k} +M_{i-1, i-1+k}
\Big)
\ee
where $1+k_\textrm{\tiny max}\leqslant i \leqslant L-k_\textrm{\tiny max}$.
We emphasise that the combinations (\ref{M-N-summation Casini}) have been found also in \cite{Arias-16} 
by employing a different approach. 
The data obtained through the combinations (\ref{M-N-summation Casini}) 
are shown in the middle panels of Fig.\,\ref{fig:sum-methods-M-Infinite} and Fig.\,\ref{fig:sum-methods-N-Infinite}
for the interval in the infinite line
and of Fig.\,\ref{fig:sum-methods-M-SemiInfinite} and Fig.\,\ref{fig:sum-methods-N-SemiInfinite}
for the interval at the beginning of the semi-infinite line.
Also in this case the agreement with the CFT predictions
(\ref{parabola-CFT-infinite}) and (\ref{parabola-CFT-semi-infinite}) 
improves as $k_{\textrm{\tiny max}}$ increases until it reaches 
an optimal value corresponding to the one adopted in the right panels.
Comparing the middle panels with the top panels and with 
Fig.\,\ref{fig:Parabola-Infinite}, Fig.\,\ref{fig:kmax-Symm-Infinite}, 
Fig.\,\ref{fig:Parabola-Semi-Infinite} and Fig.\,\ref{fig:kmax-Symm-SemiInfinite}, 
it is straightforward to notice that (\ref{M-N-summation Casini})
do not allow to describe the CFT curves close to both the endpoints of the interval. 
The data corresponding to (\ref{M-summation k^0 and k^2 der Casini}),
that are not reported in this manuscript, provide the function that vanish identically in the interval
as $L$ increases and for optimal values of $k_{\textrm{\tiny max}}$,
as expected from the CFT analysis.

%%%%%%%%%%%%%%%%%%%%%%%%%%%%%%%%%

\clearpage

\clearpage

%%%%%%%%%%%%%%%%%%%%%%%%%%%%%%%%%

Another way to obtain the results predicted by CFT in the continuum limit 
can be introduced by adapting the method employed in the numerical analysis of \cite{etp-19}
for the entanglement hamiltonian of an interval in an infinite chain of free fermions. 
Considering only blocks containing an even number $L$ of sites, 
let us decompose the operators $\widehat{H}_M$ and $\widehat{H}_N$
in (\ref{H_M and H_N operators}) respectively as
\bea
\label{H_M Eisler}
& & \hspace{-1.5cm}
\widehat{H}_M
=
L\,
\sum_{i=1}^L
\frac{M_{i,i}}{L}\,
\hat{q}_{i}^2
\\
& & \hspace{-.4cm}
\,+ 2L\,
\Bigg(\,
\sum_{k=1}^{L/2-1}
\sum_{i=1+k}^{L-k}\!
\frac{M_{i-k,i+k}}{L}\;
 \hat{q}_{i-k}\, \hat{q}_{i+k}
 \,+\!
 \sum_{k=0}^{L/2-1}\,
\sum_{i=1+k}^{L-k-1}
\frac{M_{i-k,i+k+1}}{L}\;
 \hat{q}_{i-k}\, \hat{q}_{i+k+1}
 \Bigg)
 \nonumber
\eea
and
\bea
\label{H_N Eisler}
& & \hspace{-1.5cm}
\widehat{H}_N
=
L\,
\sum_{i=1}^L
\frac{N_{i,i}}{L}\,
\hat{p}_{i}^2
\\
& & \hspace{-.4cm}
\,+ 2L\,
\Bigg(\,
\sum_{k=1}^{L/2-1}
\sum_{i=1+k}^{L-k}\!
\frac{N_{i-k,i+k}}{L}\;
 \hat{p}_{i-k}\, \hat{p}_{i+k}
 \,+\!
 \sum_{k=0}^{L/2-1}\,
\sum_{i=1+k}^{L-k-1}\!
\frac{N_{i-k,i+k+1}}{L}\;
 \hat{p}_{i-k}\, \hat{p}_{i+k+1}
 \Bigg)
 \nonumber
\eea
where we have separated the contributions of the even diagonals of $M$ and $N$ 
from the contributions of the odd diagonals.

In (\ref{H_M Eisler}) and (\ref{H_N Eisler}) the index $i$ labels the elements along the diagonals of $M$ and $N$, 
while in the decompositions introduced in \S\ref{sec:EHinHC} it corresponds to a row index. 
Treating separately the contributions of the odd diagonals and of the even diagonals in (\ref{H_M Eisler}) and (\ref{H_N Eisler})
and using (\ref{qp-field-replacement}), (\ref{q-der-field-replacement}) and (\ref{p-der-field-replacement}),
we can adapt the procedure described in \S\ref{sec:infinite-line-massless} to these decompositions of 
the operators $\widehat{H}_M$ and $\widehat{H}_N$.
In this case we find that
the combinations of diagonals occurring at the leading order as $a \to 0$ are 
\be
\label{M-summation k^0 Eisler even-odd}
 \mathsf{M}_{r_\textrm{\tiny max}}^{\textrm{e}\, (0) }(i)
 \equiv
M_{i,i} + 2 \sum_{r=1}^{r_\textrm{\tiny max}} M_{i-r,i+r}
\;\; \qquad\;\;
\mathsf{M}_{r_\textrm{\tiny max}}^{\textrm{o}\, (0) }(i)
 \equiv
M_{i,i} + 2 \sum_{r=1}^{r_\textrm{\tiny max}} M_{i-r,i+r+1}
\ee
which come from the even and odd diagonals respectively.

The next subleading order in the expansion of the entanglement hamiltonian as $a \to 0$ 
gets contributions both from $\widehat{H}_M$ and $\widehat{H}_N$.
In particular, the continuum limit of $\widehat{H}_M$ to this order gives
\be
\label{M-summation k^2 Eisler even-odd}
 \mathsf{M}_{r_\textrm{\tiny max}}^{\textrm{e}\, (2) }(i)
 \equiv
  \sum_{r=1}^{r_\textrm{\tiny max}} (2r)^2\, M_{i-r,i+r}
\;\; \qquad\;\;
\mathsf{M}_{r_\textrm{\tiny max}}^{\textrm{o}\, (2) }(i)
 \equiv
  \sum_{r=1}^{r_\textrm{\tiny max}} (2r+1)^2\, M_{i-r,i+r+1}
\ee
and the term originated from the finite difference approximation of $\mu_k''(x)$, namely
\bea
\label{M-summation k^2 der Eisler even-odd}
& & \hspace{-.8cm}
 \mathsf{M}_{2,r_\textrm{\tiny max}}^{\textrm{e}\, (2) }(i)
 \equiv
  \sum_{r=1}^{r_\textrm{\tiny max}} (2r)^2\,
  \big( M_{i-r+1,i+r+1}\,-\,
  2 M_{i-r,i+r}\, +\,
  M_{i-r-1,i+r-1}
  \big)
\\
& & \hspace{-.8cm}
\mathsf{M}_{2,r_\textrm{\tiny max}}^{\textrm{o}\, (2) }(i)
 \equiv
  \sum_{r=1}^{r_\textrm{\tiny max}} (2r+1)^2\, 
  \big( M_{i-r+1,i+r+2}\,-\,
  2 M_{i-r,i+r+1}\, +\,
  M_{i-r-1,i+r}
  \big)
    \nonumber
\eea
while from $\widehat{H}_N$ we find only
\be
\label{N-summation k^0 Eisler even-odd}
 \mathsf{N}_{r_\textrm{\tiny max}}^{\textrm{e}\, (0) }(i)
 \equiv
N_{i,i} + 2 \sum_{r=1}^{r_\textrm{\tiny max}} N_{i-r,i+r}
\;\; \qquad\;\;
\mathsf{N}_{r_\textrm{\tiny max}}^{\textrm{o}\, (0) }(i)
 \equiv
N_{i,i} + 2 \sum_{r=1}^{r_\textrm{\tiny max}} N_{i-r,i+r+1}\,.
\ee
Notice that the range of the index $i$ in the above expression is
$1+r_\textrm{\tiny max}\leqslant i\leqslant L - (1+r_\textrm{\tiny max})$
for the combinations coming from the odd diagonals 
and $r_\textrm{\tiny max} +1\leqslant i\leqslant L - r_\textrm{\tiny max}$
for the combinations coming from the even diagonals.

The final expressions providing the continuum limit of the entanglement hamiltonian
are proper combinations of the terms coming from the even and odd diagonals,
but in constructing these combinations we encounter the problem that
the former ones are defined on $L-2 r_\textrm{\tiny max}$ sites,
while in the latter ones the index $i$ assumes $L-2 r_\textrm{\tiny max}-1 $ values.
In this case the index $i$ labels the elements along the diagonals;
hence, focussing e.g. on (\ref{M-summation k^2 Eisler even-odd}),
we encounter an ambiguity in the way to combine $\mathsf{M}_{r_\textrm{\tiny max}}^{\textrm{o}\, (2) }(i) $  
with $\mathsf{M}_{r_\textrm{\tiny max}}^{\textrm{e}\, (2) }(i) $.
In the following, first we split $\mathsf{M}_{r_\textrm{\tiny max}}^{\textrm{o}\, (2) }(i)$ as $\alpha\,\mathsf{M}_{r_\textrm{\tiny max}}^{\textrm{o}\, (2) }(i)+\beta\, \mathsf{M}_{r_\textrm{\tiny max}}^{\textrm{o}\, (2) }(i)$, with $\alpha+\beta=1$. 
Then, in the sum between the even and the odd part, we choose to associate $\alpha\,\mathsf{M}_{r_\textrm{\tiny max}}^{\textrm{o}\, (2) }(i)$ to $\mathsf{M}_{r_\textrm{\tiny max}}^{\textrm{e}\, (2) }(i)$ and $\beta\,\mathsf{M}_{r_\textrm{\tiny max}}^{\textrm{o}\, (2) }(i)$ to $\mathsf{M}_{r_\textrm{\tiny max}}^{\textrm{e}\, (2) }(i+1)$.
The values of $\alpha$ and $\beta$ are not fixed uniquely and we choose  $\alpha=\beta=1/2$ in our numerical analysis.
For the interval in the infinite line, this choice guarantees the expected symmetry with respect to the center of the interval
at finite $L$.

By applying this procedure to all the expressions in 
(\ref{M-summation k^0 Eisler even-odd}), (\ref{M-summation k^2 Eisler even-odd}), 
(\ref{M-summation k^2 der Eisler even-odd}) and (\ref{N-summation k^0 Eisler even-odd}), 
we obtain respectively
\bea
\label{M-summation k^0 Eisler}
& &
\mathsf{M}_{k_\textrm{\tiny max}}^{(0)}(i)
=
\mathsf{M}_{r_\textrm{\tiny max}}^{\textrm{e}\, (0) }(i)
+
\frac{1}{2} \left(
\mathsf{M}_{r_\textrm{\tiny max}}^{\textrm{o}\, (0) }(i-1)
+
\mathsf{M}_{r_\textrm{\tiny max}}^{\textrm{o}\, (0) }(i)
\right)
\\
\rule{0pt}{.8cm}
\label{M-summation k^2 Eisler}
& &
\mathsf{M}_{k_\textrm{\tiny max}}^{(2)}(i)
=
\mathsf{M}_{r_\textrm{\tiny max}}^{\textrm{e}\, (2) }(i)
+
\frac{1}{2} \left(
\mathsf{M}_{r_\textrm{\tiny max}}^{\textrm{o}\, (2) }(i-1)
+
\mathsf{M}_{r_\textrm{\tiny max}}^{\textrm{o}\, (2) }(i)
\right)
\\
\rule{0pt}{.8cm}
\label{M-summation k^2 der Eisler}
& &
\mathsf{M}_{2,k_\textrm{\tiny max}}^{(2)}(i)
=
\mathsf{M}_{2,r_\textrm{\tiny max}}^{\textrm{e}\, (2) }(i)
+
\frac{1}{2} \left(
\mathsf{M}_{2,r_\textrm{\tiny max}}^{\textrm{o}\, (2) }(i-1)
+
\mathsf{M}_{2,r_\textrm{\tiny max}}^{\textrm{o}\, (2) }(i)
\right)
\\
\rule{0pt}{.8cm}
\label{N-summation k^0 Eisler}
& &
\mathsf{N}_{k_\textrm{\tiny max}}^{(0)}(i)
=
\mathsf{N}_{r_\textrm{\tiny max}}^{\textrm{e}\, (0) }(i)
+
\frac{1}{2} \left(
\mathsf{N}_{r_\textrm{\tiny max}}^{\textrm{o}\, (0) }(i-1)
+
\mathsf{N}_{r_\textrm{\tiny max}}^{\textrm{o}\, (0) }(i)
\right)
\eea
where $r_\textrm{\tiny max} +1\leqslant i \leqslant L-(r_\textrm{\tiny max} +1)$
and we have employed the parameter $k_\textrm{\tiny max}= 2 r_\textrm{\tiny max} + 1$
adopted throughout all this manuscript to label the last diagonal occurring in a particular combination of diagonals. 
Since we always choose odd $k_\textrm{\tiny max}$ in our numerical analysis, 
the same number of even and odd diagonals occurs in the above combinations.

The numerical results for 
(\ref{M-summation k^2 Eisler}) and (\ref{N-summation k^0 Eisler})
are reported respectively in the left and right bottom panels 
of Fig.\,\ref{fig:sum-methods-M-Infinite} and Fig.\,\ref{fig:sum-methods-N-Infinite}
for the interval in the infinite line
and of Fig.\,\ref{fig:sum-methods-M-SemiInfinite} and Fig.\,\ref{fig:sum-methods-N-SemiInfinite}
for the interval at the beginning of the semi-infinite line. 
As $L$ and $k_\textrm{\tiny max}$ increase, with $k_\textrm{\tiny max} \ll L$, 
we observe again that the agreement of the numerical data with the CFT curves 
(\ref{parabola-CFT-infinite}) and (\ref{parabola-CFT-semi-infinite}) improves. 
Nonetheless, since $r_\textrm{\tiny max} +1\leqslant i \leqslant L-(r_\textrm{\tiny max} +1)$,
the data points cannot capture the CFT curves close to the endpoints of the interval. 
We have also checked that, by performing this numerical analysis 
for (\ref{M-summation k^0 Eisler}) and (\ref{M-summation k^2 der Eisler}), 
the vanishing curve is obtained everywhere within the interval 
except in the left endpoint in the case of the semi-infinite,
where the Dirichlet boundary conditions are imposed. 
Also these results confirm the CFT predictions.

%%%%%%%%%%%%%%%%%%%%%%%%%%%%%%%%%

\clearpage

\clearpage

\end{appendices}

%%%%%%%%%%%%%%%%%%%%%%%%%%%%%%%%%%%%%%%%%%%%%%%%%

%\newpage

\end{document}